\documentclass[aps,prd,superscriptaddress,12pt,nofootinbib]{revtex4}

\usepackage{amsfonts}
\usepackage{amsmath,epsfig,bm}
\usepackage{graphicx}
\usepackage{color}

\usepackage[setpagesize=false]{hyperref}

\newcommand{\be}{\begin{equation}}
\newcommand{\ee}{\end{equation}}
\newcommand{\ba}{\begin{eqnarray}}
\newcommand{\ea}{\end{eqnarray}}
\newcommand{\bd}{\begin{displaymath}}
\newcommand{\ed}{\end{displaymath}}

\def\thalf{{\textstyle{\frac{1}{2}}}}

\def\twoth{{\textstyle{\frac{2}{3}}}}

\def\fth{{\textstyle{\frac{4}{3}}}}

\newcommand\pd{\partial}

\newcommand\la{\left\langle\,}
\newcommand\ra{\,\right\rangle}
\newcommand\vs{{v_{\rm s}}}
\newcommand\tauf{{\tau_{\rm f}}}
\newcommand\Tf{T_{\rm f}}

\newcommand\Corr{{C}}
\newcommand\pfrac[2]{\left(\frac{#1}{#2}\right)}
\newcommand\ds{d_{\rm s}}
\newcommand\cz{c_0}
\newcommand\aperp{A}

\begin{document}

\title{Relativistic Theory of Hydrodynamic Fluctuations with Applications to Heavy Ion Collisions}

\author{J. I. Kapusta}
\affiliation{School of Physics \& Astronomy, University of Minnesota, Minneapolis, MN 55455,USA}
\author{B. M\"uller}
\affiliation{Department of Physics, Duke University, Durham, NC 27708-0305, USA}
\author{M. Stephanov}
\affiliation{Department of Physics, University of Illinois, Chicago, IL 60607, USA}

\date{January 4, 2012}

\begin{abstract}

We develop the relativistic theory of hydrodynamic fluctuations for application to high energy heavy ion collisions.   In particular, we investigate their effect on the expanding boost-invariant (Bjorken) solution of the hydrodynamic
equations. We discover that correlations over a long rapidity range are induced by the propagation of the sound modes. Due to the expansion, the dispersion law for these modes is non-linear and attenuated even in the limit of zero viscosity. As a result, there is a non-dissipative wake behind the sound front which is generated by any instantaneous point-like fluctuation. We evaluate the two-particle correlators using the initial conditions and hydrodynamic
parameters relevant for heavy-ion collisions at RHIC and LHC.  In principle these correlators can be used to obtain information about the viscosities because the magnitudes of the fluctuations are directly proportional to them.

\end{abstract}

\maketitle

\section{Introduction}
\label{sec:introduction}

The success of relativistic hydrodynamics in describing the fireball created in ultrarelativistic heavy ion collisions opened the possibility to study the properties of strongly interacting matter at extremely high temperatures and densities near thermal equilibrium. We know from lattice simulations of quantum chromodynamics (QCD) that strongly interacting matter at temperatures above the crossover at $T_c \approx 165$ MeV is a quark-gluon plasma \cite{Aoki:2006br,Borsanyi:2010cj}. Lattice QCD is able to predict stationary thermodynamic properties of the 
quark-gluon plasma, such as the equation of state, but is presently unable to make reliable predictions for
dynamical properties, such as transport coefficients. 

A remarkably small value of the shear viscosity $\eta$ in the natural units of the entropy density $s$, $4\pi\eta/s < 2.5$, has been deduced from comparison of the results of relativistic viscous fluid dynamics simulations with data from Au+Au collisions at the Relativistic Heavy Ion Collider (RHIC) \cite{Song:2010mg}. This result is interesting because it is smaller by at least a factor 5 than the value of $\eta/s$ calculated in thermal perturbation theory at leading order \cite{Arnold:2003zc} and not far away from the value found earlier in a large class of strongly coupled
non-abelian gauge theories \cite{Kovtun:2004de}. It would thus be desirable to confirm the inferred experimental result for the shear viscosity by other methods.

Due to the fluctuation-dissipation theorem, the shear and bulk viscosities not only control the dissipative properties of a fluid in the limit of small velocity gradients, but they also control the magnitude of hydrodynamic fluctuations in
the fluid.\footnote{This fact is also represented by Kubo formulas,
  relating viscosities to correlators of stress-energy tensor, and
  underlies the approach taken in Ref.~\cite{Dobado:2011wc} to study
  bulk viscosity.} Thus it is interesting to explore whether fluctuations in the density and flow velocity of the fluid can be used to deduce the value of the shear viscosity from experimental data.  The purpose of our work is to lay the foundations for quantitative investigations of this idea by identifying the sensitivity of correlation observables to the hydrodynamic fluctuations.

Before proceeding, let us identify four major sources of density fluctuations in relativistic heavy ion collisions:
\begin{itemize}
\setlength{\itemsep}{0pt}
\item[(a)] {\em Initial state fluctuations:}
These are the result of quantum fluctuations in the densities of the two colliding nuclei and fluctuations of the energy deposition mechanism. They appear as event-by-event fluctuations in the energy density and flow velocity distributions at the onset of the hydrodynamic regime. These fluctuations and their phenomenological ramifications have recently been studied extensively
\cite{Shuryak:2009cy,Staig:2010pn,Staig:2011as,Staig:2011wj,Petersen:2010cw,Qin:2010pf,Schenke:2010rr,Nagle:2010zk,Qiu:2011iv,Han:2011iy,
Cheng:2011hz,Bhalerao:2011bp,Flensburg:2011wx,Schenke:2011bn} 
because they may be responsible \cite{Alver:2010gr} for the angular correlations of particle emission observed in the heavy-ion experiments \cite{Aggarwal:2010rf,Adare:2011tg,ALICE:2011vk,ATLAS:2011hf,Chatrchyan:2011eka,Gardim:2011xv}. 
The power spectrum of the final-state angular correlations induced by initial-state fluctuations may provide information about the speed of sound and the shear viscosity of the matter produced in the heavy-ion collision \cite{Staig:2011wj,Mocsy:2011xx}.
Longitudinal fluctuations and correlations among the initial-state angular fluctuations have been investigated \cite{Petersen:2011fp,Cheng:2011hz}.
 The initial state correlations over large rapidity
intervals have been subject to studies in connection with the ridge
phenomenon
\cite{Adams:2005ph,Putschke:2007mi,Voloshin:2003ud,Armesto:2004pt,Strickland:2005we,Majumder:2006wi,Shuryak:2007fu,Wong:2007mf,Dumitru:2008wn,Chiu:2008ht}.

\item[(b)] {\em Hydrodynamic fluctuations:} 
These are the result of finite particle number effects in a given fluid cell.  This leads to local thermal fluctuations of the energy density and flow velocity which propagate according to the hydrodynamical equations. According to the general theory of hydrodynamical fluctuations \cite{statphys2}, the squared amplitude of these fluctuations is proportional to the viscosity. These fluctuations are the focus of our paper.

\item[(c)] {\em Fluctuations induced by hard processes:}
Energetic partons, which have been scattered in the initial collision of the
two nuclei, can propagate through the quark-gluon plasma where they lose
energy. To the extent that this energy is thermalized, it acts as a source term
for the hydrodynamical equations. The space-time shape of this source term
has been calculated in the weak and strong coupling limit 
\cite{Neufeld:2008hs,Gubser:2007xz,Chesler:2007sv}. 
If the shear viscosity of the plasma is as low as inferred from the RHIC data,
these sources will excite Mach-cone shaped perturbations in the expanding
fluid \cite{CasalderreySolana:2004qm,Neufeld:2008dx}. It is presently not clear
whether these lead to observable phenomena after freeze-out \cite{Betz:2008wy}.

\item[(d)] {\em Freeze-out fluctuations:}
Event-by-event fluctuations may also be caused by finite particle number effects
during and after the freeze-out of the hydrodynamically expanding fluid. 
\end{itemize}

The main purpose of this article is to develop and apply the relativistic theory of hydrodynamical fluctuations to the evolution of the quark-gluon plasma formed in relativistic heavy-ion collisions.  Although the relativistic generalization of the text-book \cite{statphys2} theory of the hydrodynamic fluctuations have been considered before in a different context \cite{Salie:1983}, for completeness we outline the derivation in Sec. \ref{sec:hydro-fluct}. We then apply the resulting stochastic hydrodynamic equations to the simplest example of a boost-invariant (Bjorken) flow. Our purpose is to illustrate the application of the theory in a most transparent, yet phenomenologically meaningful setting.

We are able to obtain a number of closed form analytic results which demonstrate important phenomenological consequences of the theory. In particular, we find that hydrodynamic fluctuations during the early phase of the
expansion naturally induce correlations across large rapidity intervals.  It is usually assumed that such correlations, observed in experiments, must be produced before equilibration.

A full implementation of the framework we develop here will eventually require numerical solution of stochastic hydrodynamic equations in three spatial dimensions, which we defer to future studies.
 
\section{Theory of Hydrodynamic Fluctuations}
\label{sec:hydro-fluct}

Hydrodynamics is  an effective theory that describes the long wavelength and low frequency space-time evolution of the densities of a few conserved quantities such as energy, momentum, electric charge and baryon number.  In the case of spontaneous breaking of continuous symmetries it also describes the evolution of the phases of order parameters. These hydrodynamic variables are defined as average values of the corresponding local, space-time dependent, coarse-grained operators.  The coarse-grained averaging is performed over distances and times that are small compared to the macroscopic scales of interest but large compared to the microscopic scales, such as mean free paths and mean free times between collisions. The hydrodynamic variables evolve according to the deterministic equations which follow from the conservation equations obeyed by the corresponding operators.

Fluctuations and correlations in the hydrodynamic variables can be characterized by averaged values of the products of the operators at different space-time points.  Although the fluctuations themselves occur on microscopically short space-time scales, these fluctuations are correlated not only on short space-time scales but also on macroscopically large scales. This can be understood as a result of the diffusion or propagation of each fluctuation at any earlier time to later times. Such propagation over long times and distances, and thus the long-range behavior of correlation functions, is described by hydrodynamics.

\subsection{Hydrodynamic variables and equations}

We consider a general case of a system with 5 conserved quantities: energy, charge and three momentum components. In the case of QCD we can think of the charge being the baryon number.  The five equations are the conservation equations for the energy-momentum tensor, $\pd_\mu T^{\mu\nu}=0$, and the current
conservation equation $\pd_\mu J_B^\mu=0$.

The energy-momentum tensor and current densities for a fluid in thermal, chemical and
mechanical equilibrium are
\be
T^{\mu\nu}_{\rm ideal} = - Pg^{\mu\nu}+wu^{\mu}u^{\nu}
;\qquad J_{\rm B}^\mu= n_{\rm B}u^\mu
\qquad\mbox{(equilibrium)}
\ee
Here $P$ is the equilibrium pressure at given energy density $\epsilon$ and baryon density $n_B$,   $w=P+\epsilon=Ts + \mu_B n_B$ is the local enthalpy density, $\mu_B$ is the baryon chemical potential
and $u^{\mu}$ is the local flow 4-velocity.  The metric $g_{\mu\nu}$ is $(+,-,-,-)$.  

The non-equilibrium corrections to these expressions, $\Delta T^{\mu\nu}$ and $\Delta J_{\rm B}^{\mu}$, are proportional, at lowest order, to first derivatives of the local quantities with coefficients given by the shear viscosity $\eta$, bulk viscosity $\zeta$, and thermal conductivity $\chi$.  Explicit expressions may be found in textbooks \cite{fluid1,fluid2}.  Local thermal fluctuations are described by the additional stochastic terms $S^{\mu\nu}$ and $I^{\mu\nu}$.
\begin{align}
  T^{\mu\nu} &= T^{\mu\nu}_{\rm ideal} + \Delta T^{\mu\nu} +
  S^{\mu\nu}
  \nonumber \\
  J_{\rm B}^{\mu} &= n_{\rm B} u^{\mu} + \Delta J_{\rm B}^{\mu} +
  I^{\mu\nu}
\end{align}
In the following we shall determine the correlation functions of the stochastic terms. Since the source of the fluctuations is local, these correlation functions are delta-functions in space and time. The amplitude of these source terms is fixed by the fluctuation-dissipation theorem.

In practice, the idea is to consider the stochastic terms as given functions of space and time and to solve the fluid equations of motion to first order in it.  Quantities which are linear in the $S^{\mu\nu}$ will average to zero, where the average is taken over the ensemble of fluctuations.  Quantities which are quadratic in the $S^{\mu\nu}$ may have non-zero average values which we must determine.

The form of the hydrodynamic equations depend on the definition of the local flow velocity $u^\mu$. There are two common choices; we discuss each in turn, including the modifications necessary to incorporate fluctuations.

\subsection{Eckart approach}

The Eckart approach is a convenient choice if we want to compare with the non-relativistic limit.  In this approach $u^{\mu}$ is the velocity of baryon number flow.  The dissipative terms must satisfy the conditions
$\Delta J^{\mu}_{\rm B}=0$ and $u_{\mu} u_{\nu} \Delta T^{\mu\nu} = 0$, the latter following from the 
requirement that $T^{00}$ be the energy density in the local (baryon) rest frame.  The most general form of $\Delta T^{\mu\nu}$ is
\be
\Delta T^{\mu\nu} = \Delta T^{\mu\nu}_{\rm vis} + \Delta T^{\mu\nu}_{\rm heat}
\ee
where
\be
\Delta T^{\mu\nu}_{\rm vis} = 
\eta \left(\Delta^{\mu} u^{\nu} + \Delta^{\nu} u^{\mu}\right)
+\left(\twoth \eta - \zeta\right) h^{\mu\nu} \left( \partial \cdot u \right)
\ee
is the viscous part and
\be
\Delta T^{\mu\nu}_{\rm heat} = 
\chi \left( h^{\mu\alpha} u^{\nu} + h^{\nu\alpha} u^{\mu}
\right) q_{\alpha}
\ee
is the heat conduction part.  Here
\be
h^{\mu\nu} = u^{\mu} u^{\nu} - g^{\mu\nu}
\label{defineh}
\ee
is a projection tensor normal to $u^{\mu}$,
\be
\Delta_{\mu} = \partial_{\mu} - u_{\mu} \left( u \cdot \partial \right)
\ee
is a derivative normal to $u^{\mu}$, and
\be
q_{\alpha} = -\partial_{\alpha} T + T \left( u \cdot \partial \right) u_{\alpha}
\ee
is a four-vector whose nonrelativistic limit is ${\bf q} = 
\mbox{\boldmath $\nabla$} T$.  The entropy current is
\be
s^{\mu} = s u^{\mu} + \frac{1}{T} u_{\nu} \Delta T^{\mu\nu} \, .
\ee
By using energy-momentum conservation $\partial_{\mu} T^{\mu\nu}=0$ the divergence of the entropy current can be put in the compact form
\be
\partial_{\mu}s^{\mu} = \Delta T^{\mu\nu} \partial_{\mu} \left( \beta u_{\nu} \right) \, .
\ee
For some purposes it is better to express this divergence as
\be
\partial_{\mu}s^{\mu} = \Delta T^{\mu\nu} 
\left[\frac{1}{2T} \left( \Delta_{\mu} u_{\nu} + \Delta_{\nu} u_{\mu} \right)
+ \frac{1}{2T^2} \left( u_{\mu} q_{\nu} + u_{\nu} q_{\mu} \right) \right]
\, .
\ee
Substituting the explicit form of $\Delta T^{\mu\nu}$ into the above expression gives
\ba
\partial_{\mu}s^{\mu} &=& \frac{\eta}{2T}
\left[\left( \Delta^{\mu} u^{\nu} + \Delta^{\nu} u^{\mu} \right)
+ \twoth h^{\mu\nu} \left( \partial \cdot u \right) \right]^2 \nonumber \\
& & \mbox{} + \frac{\zeta}{T} \left( \partial \cdot u \right)^2
+ \frac{\chi}{T^2} h^{\mu\nu} q_{\mu} q_{\nu} \, .
\ea
In the local rest frame this is
\ba
\partial_{\mu}s^{\mu} &=& \frac{\eta}{2T}
\left( \partial_iu^j + \partial_ju^i - \twoth \delta^{ij} \nabla \cdot {\bm u}
\right)^2 \nonumber \\
& & \mbox{} + \frac{\zeta}{T} \left( \nabla \cdot {\bm u} \right)^2
+ \frac{\chi}{T^2} \left( {\bm\nabla} T + T \dot{{\bm u}}\right)^2 \, .
\ea
The term $T \dot{{\bm u}}$ is a relativistic correction to $\bm\nabla T$, being smaller by a factor of $1/c^2$ in physical units.  All three dissipation coefficients must be non-negative to insure that entropy can never decrease.

It is useful to decompose $S^{\mu\nu}$ into a piece associated with viscosity and another piece associated with heat conduction.  Overall we must require that $u_{\mu} u_{\nu} S^{\mu\nu}=0$, just like $\Delta T^{\mu\nu}$, so that in the local rest frame of the fluid $T^{00}$ equals $\epsilon=T^{00}_{\rm ideal}$.  Then if we are given $S^{\mu\nu}$ with this property we can define
\be
S^{\mu\nu}_{\rm heat} = S^{\mu\alpha} u_{\alpha} u^{\nu} +
S^{\nu\alpha} u_{\alpha} u^{\mu}
\ee
and
\be
S^{\mu\nu}_{\rm vis} = S^{\mu\nu} - S^{\mu\nu}_{\rm heat} \, .
\ee
This decomposition is unique.

We follow Section 88 of \cite{statphys2} on hydrodynamic fluctuations.  In the general theory of quasi-stationary fluctuations, presented in \cite{statphys1}, one considers the set of equations
\be
\dot{x}_a = - \sum_b \gamma_{ab} X_b + y_a
\ee
which gives the response of the set of variables $x_a$ to the driving terms $X_a$ and to the $y_a$, which represent random fluctuations.  The time rate of change of the entropy is
\be
\dot{S} = - \sum_a \dot{x}_a X_a \, .
\ee
In order for the probability distribution of fluctuating variables to agree with the thermodynamic
distribution given by $e^S$, the noise autocorrelations must be given by
\be
\langle y_a(t_1) y_b(t_2) \rangle = (\gamma_{ab} + \gamma_{ba} ) \delta(t_1-t_2) \, .
\ee
This  general framework needs to be applied to the present situation.

The time rate of change of the total entropy of the system is
\be
\frac{dS}{dt} = \int d^3x \, \Delta T^{\mu\nu} 
\left[\frac{1}{2T} \left( \Delta_{\mu} u_{\nu} + \Delta_{\nu} u_{\mu} \right)
+ \frac{1}{2T^2} \left( u_{\mu} q_{\nu} + u_{\nu} q_{\mu} \right) \right] \, .
\ee
Coarse graining is performed in the usual way with cell volumes $\Delta V$.  Since viscosity and heat conduction are independent physical processes, it is natural to make the identifications
\ba
\dot{x}_1 & \rightarrow & \Delta T^{\mu\nu}_{\rm vis} \, , \nonumber \\
\dot{x}_2 & \rightarrow & \Delta T^{\mu\nu}_{\rm heat} \, .
\ea
Comparing to the rate of entropy change allows us to deduce that
\ba
X_1 & \rightarrow & -\frac{1}{2T} \left[ \Delta_{\mu} u_{\nu} + 
\Delta_{\nu} u_{\mu} \right] \Delta V \, , \nonumber \\
X_2 & \rightarrow &  
- \left[\frac{1}{2T} \left( \Delta_{\mu} u_{\nu} + \Delta_{\nu} u_{\mu} \right)
+ \frac{1}{2T^2} \left( u_{\mu} q_{\nu} + u_{\nu} q_{\mu} \right) \right]
\Delta V \, .
\ea
Next the Onsager coefficients $\gamma_{ab}$ can be determined.
\ba
\gamma_{11} &=& 2T \left[ \eta h^{\mu\alpha} h^{\nu\beta} + \thalf
\left(\zeta - \twoth \eta \right) h^{\mu\nu} h^{\alpha\beta} \right]
\frac{1}{\Delta V} \, , \nonumber \\
\gamma_{22} &=& 2 \chi T^2 \left[ h^{\mu\alpha} u^{\nu} u^{\beta}
+ h^{\nu\beta} u^{\mu} u^{\alpha} \right] \frac{1}{\Delta V} \, .
\ea
The $\gamma_{11}$ is made unique by the requirement that it vanish when any of its indices is contracted with the four-velocity.  The $\gamma_{12}$ and $\gamma_{21}$ are zero as expected.

Different coarse grained cells are independent.  Then the factor $1/\Delta V$ goes over to a Dirac delta function in position space.  The correlation functions are easily written down (after acknowledgement that they must have certain symmetries in the Lorentz indices).  They are
\bd
\langle S^{\mu\nu}_{\rm vis}(x_1) S^{\alpha\beta}_{\rm vis}(x_2) \rangle =
2T \left[ \eta \left( h^{\mu\alpha} h^{\nu\beta} + h^{\mu\beta} h^{\nu\alpha}
\right) \right.
\ed
\be
+ \left. \left(\zeta - \twoth \eta \right) h^{\mu\nu} h^{\alpha\beta} \right]
\delta (x_1 - x_2)
\label{eq:Eckartvis}
\ee
and
\bd
\langle S^{\mu\nu}_{\rm heat}(x_1) S^{\alpha\beta}_{\rm heat}(x_2) \rangle =
2\chi T^2 \left[ h^{\mu\alpha} u^{\nu} u^{\beta} + 
h^{\nu\beta} u^{\mu} u^{\alpha} \right.
\ed
\be
+ \left. h^{\mu\beta} u^{\nu} u^{\alpha} + h^{\nu\alpha} 
u^{\mu} u^{\beta}\right] \delta (x_1 - x_2)
\label{eq:Eckartheat}
\ee
and
\be
\langle S^{\mu\nu}_{\rm vis}(x_1) S^{\alpha\beta}_{\rm heat}(x_2) \rangle = 0
\, .
\ee
When the viscous correlation function is evaluated in the local rest frame it will vanish unless all of the indices are spatial.  With $\mu\nu = ik$ and $\alpha\beta = lm$ we get
\be
\langle S^{ik}_{\rm vis}(x_1) S^{lm}_{\rm vis}(x_2) \rangle =
2T \left[ \eta \left( \delta_{il} \delta_{km} + \delta_{im} \delta_{kl}
\right) 
+ \left(\zeta - \twoth \eta \right) \delta_{ik} \delta_{lm} \right]
\delta (x_1 - x_2)
\ee 
which is exactly the expression in \cite{statphys2}.  When the heat correlation function is evaluated in the local rest frame it will vanish unless each $S_{\rm heat}$ has one spatial and one temporal index.  With $\mu\nu = 0i$ and $\alpha\beta = 0j$ we get
\be
\langle S^{0i}_{\rm heat}(x_1) S^{0j}_{\rm heat}(x_2) \rangle =
2\chi T^2 \delta_{ij} \delta (x_1 - x_2)
\ee
which also agrees with the corresponding expression in \cite{statphys2}.  Since these correlation functions reduce to the known ones in the local rest frame, and since they are constructed from tensors, they are obviously valid in any frame of reference.

\subsection{Landau-Lifshitz approach}

The Landau-Lifshitz approach is the most convenient and frequently used approach for ultrarelativistic heavy ion collisions.  In this approach $u^{\mu}$ is the velocity of energy transport.  
The dissipative part of the energy-momentum tensor satisfies $u_{\mu} \Delta 
T^{\mu\nu} = 0$, and $\Delta J_{\rm B}^{\mu}$ is not constrained to be zero.
In this case the most general form of the energy-momentum tensor is
\be
\Delta T^{\mu\nu} = 
\Delta T^{\mu\nu}_{\rm vis} = 
\eta \left(\Delta^{\mu} u^{\nu} + \Delta^{\nu} u^{\mu}\right)
+\left(\twoth \eta - \zeta\right) h^{\mu\nu} \partial \cdot u \, .
\label{LLvis}
\ee
The baryon current is modified by
\be
\Delta J^{\mu}_{\rm B} = \sigma T \Delta^{\mu} \left(\beta \mu_B \right) \, ,
\ee
where $\sigma$ is the (baryon) charge conductivity.  The modification to the current satisfies 
$u_{\mu} \Delta J^{\mu}_{\rm B} = 0$.  This means that $n_{\rm B}$ is the baryon density in the local rest frame.  

The entropy current in this approach is different, being
\be
s^{\mu} = s u^{\mu} - \beta \mu_{\rm B} \Delta J_{\rm B}^{\mu} \, .
\ee
Using baryon number conservation, $\partial_{\mu} J^{\mu}_{\rm B} = 0$, we can write
\be
\partial_{\mu} s^{\mu} = \partial_{\mu}\left( s u^{\mu} \right)
+ \beta \mu_{\rm B} \partial_{\mu} \left( n_{\rm B} u^{\mu} \right) - \Delta J_{\rm B}^{\mu} \partial_{\mu}\left( \beta \mu_{\rm B} \right)
\, .
\ee
By using energy-momentum conservation in the form $u_{\mu} \partial_{\nu}
T^{\mu \nu} = 0$, this can be written in a way convenient for future use.
\be
\partial_{\mu}s^{\mu} = \Delta T^{\mu\nu}_{\rm vis} 
\left[\frac{1}{2T} \left( \Delta_{\mu} u_{\nu} + \Delta_{\nu} u_{\mu} \right)
\right] +  \Delta J_{\rm B}^{\mu} \left[ h_{\mu \nu} \Delta^{\nu} \left(
\beta \mu_{\rm B} \right) \right]
\, .
\label{LLproduce}
\ee
Compared to the Eckart frame there is no change in the viscous part associated with shear and bulk
viscosities.  Therefore it can again be written in the symmetric form
\ba
\partial_{\mu}s^{\mu} &=& \frac{\eta}{2T}
\left[\left( \Delta^{\mu} u^{\nu} + \Delta^{\nu} u^{\mu} \right)
+ \twoth h^{\mu\nu} \left( \partial \cdot u \right) \right]^2 
+ \frac{\zeta}{T} \left( \partial \cdot u \right)^2 \nonumber \\
& & \mbox{} 
+ \sigma T 
h^{\mu \nu} 
\Delta_{\mu} \left( \beta \mu_{\rm B} \right)
\Delta_{\nu} \left( \beta \mu_{\rm B} \right) \, .
\ea

The part due to charge conductivity seems to be different than the part due to heat conduction in the Eckart frame, but it is not.  Using energy-momentum conservation in the form 
$h_{\alpha\mu } \partial_{\nu} T^{\mu\nu}_{\rm ideal}= 0$, which is valid to zeroth order in the dissipative coefficients and sufficient for this purpose, and $dP=s dT+ n_{\rm B}d\mu_{\rm B}$, one finds
\begin{equation}
  \label{eq:beta-mu-q}
  \Delta_\alpha(\beta\mu_{\rm B}) =  \frac{w}{n_{\rm B}T^2}q_\alpha \, .
\end{equation}
This can be inserted into the expression for the divergence of the entropy current to obtain
exactly the same expression as in the Eckart frame, provided that the charge conductivity $\sigma$ is related to the heat conductivity $\chi$, by
\begin{equation}
\sigma
=\chi T(n_{\rm B}/w)^2,\label{eq:sigma-chi}
\end{equation}
which corresponds to the Franz-Wiedemann law.  

The fluctuations $S^\mu=S^{\mu\nu}_{\rm vis}$ and $I^\mu$ must satisfy the conditions $u_{\mu} S^{\mu\nu}_{\rm vis} = 0$ and $u_{\mu} I^{\mu} = 0$ for the reasons mentioned above.

The time rate of change of the total entropy of the system is
\be
\frac{dS}{dt} = \int d^3x \left\{ \Delta T^{\mu\nu}_{\rm vis} 
\left[\frac{1}{2T} \left( \Delta_{\mu} u_{\nu} + \Delta_{\nu} u_{\mu} \right)
\right] +  \Delta J_{\rm B}^{\mu} \left[ h_{\mu \nu} \Delta^{\nu} \left(
\beta \mu_{\rm B} \right) \right] \right\} \, .
\ee
It is natural to make the identifications
\ba
\dot{x}_1 & \rightarrow & \Delta T^{\mu\nu}_{\rm vis} \, , \nonumber \\
\dot{x}_2 & \rightarrow & \Delta J^{\mu}_{\rm B} \, .
\ea
Comparing to the rate of entropy change allows us to deduce that
\ba
X_1 & \rightarrow & -\frac{1}{2T} \left[ \Delta_{\mu} u_{\nu} + 
\Delta_{\nu} u_{\mu} \right] \Delta V \, , \nonumber \\
X_2 & \rightarrow &  -h_{\mu\nu} \Delta^{\nu} 
\left( \beta \mu_{\rm B} \right) \Delta V \, .
\ea
Next the $\gamma_{ab}$ can be determined.
\ba
\gamma_{11} &=& 2T \left[ \eta h^{\mu\alpha} h^{\nu\beta} + \thalf
\left(\zeta - \twoth \eta \right) h^{\mu\nu} h^{\alpha\beta} \right]
\frac{1}{\Delta V} \, , \nonumber \\
\gamma_{22} &=& \sigma T 
h^{\mu\nu} \frac{1}{\Delta V} \, .
\ea
The $\gamma_{11}$ is made unique by the requirement that it vanish when any of its indices is contracted with the four-velocity.  The $\gamma_{22}$ is made unique by the requirement that $u_{\mu} \Delta J^{\mu}_{\rm B} = 0$.  The $\gamma_{12}$ and $\gamma_{21}$ are again zero as expected.

The correlation function for the viscous part
$\langle S^{\mu\nu}_{\rm vis}(x_1) S^{\alpha\beta}_{\rm vis}(x_2) \rangle$
is exactly the same as in the Eckart approach.  The mixture
$\langle S^{\mu\nu}_{\rm vis}(x_1) I^{\alpha}(x_2) \rangle$ is zero.  The correlation function for the baryon current is
\be
\langle I^{\mu}(x_1) I^{\nu}(x_2) \rangle = 2 \sigma T
h^{\mu\nu} \delta (x_1 - x_2) \, .
\ee
When the baryon current correlation function is evaluated in the local rest frame it will vanish unless both indices are spatial.  Then
\be
\langle I^i(x_1) I^j(x_2) \rangle = 2 \sigma T
\delta_{ij} \delta (x_1 - x_2) \, .
\ee 
This completes the generalization of the theory of hydrodynamic fluctuations to the relativistic domain.

\section{Fluctuations in Boost Invariant Hydrodynamics}
\label{sec:an-example}

In this section we consider, as an example, application of the stochastic hydrodynamic equations derived in the previous section to the hydrodynamic fluctuations around Bjorken's boost-invariant solution of relativistic hydrodynamics \cite{Bjorken:1982qr}. Unlike the thermal fluctuations around a stationary equilibrium solution, which are well-known, the correlations induced by hydrodynamic fluctuations on a non-stationary solution have not been discussed in the literature to our knowledge.  Although this example is not entirely realistic or directly applicable to data, it is semi-analytic in nature and allows us to gain experience with these fluctuations and with the issues that may arise in more realistic, multi-dimensional calculations.

Here we shall consider only fluctuations of temperature (or energy density) and flow velocity and neglect the effects of the baryon number fluctuations. For highly relativistic heavy ion collisions at LHC and the top range of RHIC energies this is a reasonable approximation because the smallness of the baryon chemical potential $\mu_{\rm B}$ suppresses mixing between baryon charge and energy density fluctuations.

In this example we shall focus on longitudinal flow fluctuations by integrating all densities over the coordinates $x$ and $y$ perpendicular to the beam or $z$ axis. This effectively reduces the dimensionality of the problem to $(1+1)$.
Thus, our example is different from the treatment in the existing literature in at least in two aspects: (i) we consider hydrodynamic fluctuations, not initial state fluctuations; and (ii) we consider longitudinal correlations, not
azimuthal ones. We shall briefly discuss transverse correlations in Appendix \ref{sec:azim} but defer their detailed study to further work.

It is convenient to view the Bjorken boost-invariant flow in Bjorken coordinates: proper time $\tau$ and spatial rapidity $\xi$.
\ba
\tau &=& \sqrt{t^2-z^2} \nonumber \\
\xi &=& \tanh^{-1}(z/t) \nonumber \\
t &=& \tau \cosh \xi \nonumber \\
z &=& \tau \sinh \xi
\ea
The average values of hydrodynamic quantities depend only on $\tau$ while fluctuations, after integration over the transverse coordinates $x$ and $y$, depend on both, $\tau$ and $\xi$.  The flow velocity is given by $u^\mu=x^\mu/\tau + \delta u^\mu$,  where the last term denotes the fluctuations.  We express the fluctuations of the longitudinal flow in terms of the rapidity variable $\omega$ which we define as
\ba
u^0 &=& \cosh (\xi+\omega(\xi,\tau)) \nonumber \\
u^3 &=& \sinh (\xi+\omega(\xi,\tau)) \, .
\ea
The local pressure depends on the temperature which in turn depends on both coordinates.  The average value of $T$ depends only on the proper time, but fluctuations of $T$ depend on both coordinates.  Therefore
\ba
T &=& T_0(\tau) + \delta T(\xi,\tau) \nonumber \\
P &=& P_0(\tau) + \delta P(\xi,\tau) \nonumber \\
\epsilon &=& \epsilon_0(\tau) + \delta \epsilon(\xi,\tau) \, ,
\ea
where the subscript 0 refers to the average value of the function.  Obviously all variations are related to variations in the temperature on account of the equations of state.
\ba
\delta \epsilon &=& c_V(T) \delta T \nonumber \\
\delta s &=& \frac{c_V(T)}{T} \delta T \nonumber \\
\delta P &=& s(T) \delta T \nonumber \\ 
\delta w &=& \delta \epsilon+\delta P \, .
\label{dTs}
\ea
Here $c_V(T) = d\epsilon/dT$ is the heat capacity per unit volume.  For the case of zero chemical potentials, as we are considering here, $c_V(T)=s(T)/{\vs}^2(T)$.
 
The noise term satisfies $u_{\mu} S^{\mu\nu}=0$ and is symmetric in its indices.  Due to the reduced (1+1) dimensionality of this model this condition allows us to express the noise in terms of a single scalar function $f$ as
\be
S^{\mu\nu}=w(\tau) f(\xi,\tau) h^{\mu\nu}
\label{hS}
\ee
where $h^{\mu\nu}$ was defined in Eq. (\ref{defineh}); the factor of $w(\tau)$ is included to make $f$ dimensionless and to simplify subsequent formulas.  For the same reason the viscous term can be expressed in terms of a single function
\be
\Delta T^{\mu\nu}_{\rm vis} = - \left( \fth \eta + \zeta \right) (\partial \cdot u) h^{\mu\nu} \, .
\label{hT}
\ee
The fluctuations mentioned above will be linear functionals of $f$.  Their average values will be zero since $\langle f \rangle = 0$.  The fluctuations in those observables will be determined by
\begin{equation}\label{eq:ff-xi}
\langle f(\xi_1,\tau_1)  f(\xi_2,\tau_2) \rangle =
\frac{2T(\tau_1)}{A \tau_1 w^2(\tau_1)} \left[ \fth \eta(\tau_1) + \zeta(\tau_1) \right]
\delta\left(\tau_1-\tau_2\right) \delta\left(\xi_1-\xi_2\right)
\end{equation}
on account of Eq. (\ref{eq:Eckartvis}).  In this case the delta-function in the transverse coordinates 
$\delta\left({\bf x}_{\perp 1}-{\bf x}_{\perp 2}\right)$ is replaced with $1/A$, where $A$ is the effective transverse area of the colliding nuclei (for noncentral collisions it would depend on the impact parameter in the usual way).

\subsection{Hydrodynamic equations}
\label{sec:hydr-equat}

The hydrodynamic equations of motion can now be written out using any one of several standard methods.  There are two independent scalar
equations of motion each of which is first order in derivatives.  In the absence of fluctuations, one of them is satisfied automatically
due to the assumption of boost invariance.  When dissipation is neglected the nontrivial equation simply expresses entropy conservation for ideal fluid flow
\begin{equation}
\frac{d(\tau s)}{d\tau} = 0
\label{eq:dts/dt}
\end{equation}
and has the solution $s(\tau)=s(\tau_0)\tau_0/\tau$ where $\tau_0$ is the initial proper time (when thermalization first is achieved). Once dissipation is included the equation becomes more complicated.
Defining
\begin{equation}
\nu \equiv (4\eta/3+\zeta)/s,\label{eq:nu-def}
\end{equation}
the equation is
\begin{equation}
\frac{d(\tau s)}{d\tau} = \frac{\nu s}{\tau T}
\label{eq:d-tau-s}
\end{equation}
meaning that the entropy per unit rapidity interval, $\tau s A$, increases due to dissipation.  The explicit solution requires knowing the relationship between $s$ and $T$, in other words the equation of state, plus the temperature dependence of $\nu$.  For example, using an equation of state with ${\vs}^2 = dP/d\epsilon =$ constant, and with $\nu =$ constant, the solution is
\begin{equation}
T(\tau) = \left[ T_0 +  \frac{\vs^2 \gamma_s^2 \nu}{\tau_0} \right]
\left(\frac{\tau_0}{\tau}\right)^{\vs^2} -  \frac{\vs^2 \gamma_s^2\nu}{\tau}
\label{eq:T-nu-solution}
\end{equation}
where $\gamma_s=1/\sqrt{1-{\vs}^2}$.  Compared to the inviscid case the temperature decreases more slowly, assuming it starts from the same value. 

Now we account for fluctuations by adding noise.  At this point, we make no assumption about the form of the equation of state or the temperature dependence of $\nu$.  The two independent equations that follow are
\begin{equation}
\tau \frac{\partial \delta \epsilon}{\partial \tau} + \delta w  +wf -  \frac{\delta (\nu s)}{\tau} + \left[ w - 2 \frac{\nu s}{\tau} \right] \frac{\partial \omega}{\partial \xi}  = 0
\label{11eq}
\end{equation}
and
\begin{equation}
\tau \frac{\partial}{\partial \tau} \left[ \omega \left( w - \frac{\nu s}{\tau} \right) \right] 
 + 2 \omega \left( w - \frac{\nu s}{\tau} \right) 
+ \frac{\partial}{\partial \xi} \left[ \delta P + wf - \frac{\delta (\nu s)}{\tau} \right]
-\frac{\nu s}{\tau} \frac{\partial^2 \omega}{\partial \xi^2} =0 \, .
\label{22eq}
\end{equation}
In deriving these equations it is helpful to make use of Eqs. (\ref{hS}) and (\ref{hT}).  On account of the reduced dimensionality, as reflected in (\ref{hS}) and (\ref{hT}), Eqs. (\ref{11eq}) and (\ref{22eq}) follow from the equations of motion of a perfect fluid ($f=0$ and $\nu=0$) by the replacements
\bd
P \rightarrow P + wf - \frac{\nu s}{\tau} \left( 1 + \frac{\partial \omega}{\partial \xi} \right)
\ed
while $\delta \epsilon$ is unchanged.  The fluctuations $\delta P$, $\delta s$, $\delta \epsilon$, and $\delta w$ can all be expressed in terms of a new dimensionless variable
\begin{equation}
 \rho \equiv \delta s/s,\label{eq:rho-def}
\end{equation}
so that $\delta\epsilon=w \rho$ and $\delta P=\vs^2 w \rho$. Hence the pair of equations (\ref{11eq}) and (\ref{22eq}) will determine the two independent dimensionless variables $\rho$ and $\omega$.

Since the unperturbed solution is boost-invariant and independent of $\xi$, it is advantageous to use the Fourier transform
\begin{equation}\label{eq:FT}
\tilde{X}(k,\tau) = \int_{-\infty}^{\infty} d\xi {\rm e}^{-ik\xi} X(\xi,\tau) 
\end{equation}
for any variable $X$.  Note that the wavenumber $k$ is dimensionless.  With this transformation Eqs. (\ref{11eq}) and (\ref{22eq}) become a pair of coupled first order linear differential equations.  The solutions for the dimensionless variables can be expressed as
\begin{equation}
\tilde{\rho}(k,\tau) = - \int_{\tau_0}^{\tau} \frac{d\tau'}{\tau'}
\tilde{G}_{\rho}(k;\tau,\tau') \tilde{f}(k,\tau')\label{eq:rho-G-f}
\end{equation}
and
\begin{equation}
\tilde{\omega}(k,\tau) = - \int_{\tau_0}^{\tau} \frac{d\tau'}{\tau'}
\tilde{G}_{\omega}(k;\tau,\tau') \tilde{f}(k,\tau')\label{eq:omega-G-f}
\end{equation}
Note that $\tilde{\rho}(k,\tau_0)=0$ and $\tilde{\omega}(k,\tau_0)=0$
so that there are no fluctuations in the initial conditions, although
they could easily be incorporated (see Sec. \ref{sec:contr-init-state}).

The problem reduces to finding the (dimensionless) Green functions $G_{\rho}$ and $G_{\omega}$ followed by quadrature.  Averaging is performed by use of
\be
\langle \tilde{f}(k_1,\tau_1)  \tilde{f}(k_2,\tau_2) \rangle =
\frac{2\,\nu(\tau_1)}{A \tau_1 w(\tau_1)} \delta\left(\tau_1-\tau_2\right) 2\pi\, \delta\left(k_1+k_2\right) \, .
\label{k-correlation}
\ee
In $k$-space the correlators are
\begin{equation}
  \label{eq:XY-k}
  \la \tilde{X}(k_1,\tau_1) \, \tilde{Y}(k_2,\tau_2) \ra
= \frac{2\pi}{A} \delta(k_1 + k_2) \int\limits_{\tau_0}^{\min(\tau_1,\tau_2)} \frac{d\tau}{\tau^3} \frac{2\,\nu(\tau)}{w(\tau)}
 \, \tilde{G}_{X}(k_1;\tau_1, \tau) \tilde{G}_{Y}(k_2;\tau_2, \tau)
\end{equation}
where $X$ and $Y$ can be either $\rho$ or $\omega$. The correlator in  $\xi$-space is obtained by a Fourier transform (\ref{eq:FT}).

For the most part we shall be interested in the equal-(proper)time correlation function at the freeze-out time $\tauf$, which can be written as
\begin{equation}
  \label{eq:Corr-def}
\Corr_{XY}(\xi_1-\xi_2;\tauf) \equiv  \la X(\xi_1,\tauf) \, Y(\xi_2,\tauf) \ra
= \frac{2}{A} \int\limits_{\tau_0}^\tauf
\frac{d\tau}{\tau^3} \frac{\nu(\tau)}{w(\tau)}
\,G_{XY}(\xi_1-\xi_2;{\tauf},\tau)\,,
\end{equation}
where the Fourier transform of $G_{XY}(\xi;\tauf,\tau)$ is given by
\begin{equation}
  \label{eq:G-GG}
\tilde{G}_{XY}(k;\tauf,\tau) \equiv \tilde{G}_{X}(k;\tauf, \tau) \tilde{G}_{Y}(-k;\tauf, \tau) \, .
\end{equation}
Thus
\begin{equation}
  \label{eq:G-GG-xi}
{G}_{XY}(\xi_1-\xi_2;\tauf,\tau) =\int_{-\infty}^\infty d\xi{G}_{X}(\xi_1-\xi;\tauf, \tau) {G}_{Y}(\xi_2-\xi;\tauf, \tau) \, .
\end{equation}
This equation shows directly that a fluctuation at point $\xi$ at time $\tau$ induces a correlation between points $\xi_1$ and $\xi_2$ at later time
$\tauf$ via a hydrodynamically propagating response given by Eqs. (\ref{eq:rho-G-f}) and (\ref{eq:omega-G-f}).

\subsection{Inviscid case}

For the sake of clarity we shall first present the case where we neglect the contribution of dissipation (viscosity) in the equations of motion and then later consider viscous corrections.  Of course we cannot simply set the viscosity to zero since, according to the fluctuation-dissipation theorem expressed by Eq. (\ref{eq:Eckartvis}), we would also eliminate fluctuations. However, if the viscosity is small and the flow is close to ideal, as is the case for heavy ion collisions, the contribution of viscous terms to the correlators is limited to the vicinity of singularities. These singularities correspond to unattenuated propagation of sound shocks which the viscosity will smear out, as we shall quantify in Sec. \ref{sec:viscosity}.

After Fourier transformation, Eqs. (\ref{11eq}) and (\ref{22eq}) become
\begin{equation}
\tau \frac{\partial \tilde{\rho}}{\partial \tau} +ik\tilde{\omega} + \tilde{f} = 0
\label{11eqF}
\end{equation}
\begin{equation}
\tau \frac{\partial \tilde{\omega}}{\partial \tau} + \left( 1 - {\vs}^2\right) \tilde{\omega} +ik {\vs}^2 \tilde{\rho} +ik \tilde{f}  = 0
\label{22eqF}
\end{equation}
There are at least two different methods to solve these equations, each having their own merits.  They must, of course, yield the same solution.  We outline each in turn.

The pair of equations can be combined into a Langevin equation for the two-component vector
\be
\label{eq:psi-def}
{\tilde{\bm \psi}}=\begin{pmatrix}
\tilde{\rho} \\ \tilde{\omega}
\end{pmatrix} .
\ee 
The Langevin equation takes the form
\begin{equation}\label{eq:dtpsi}
\tau \frac{\partial \tilde{{\bm \psi}}}{\partial \tau} + {\bm D} \tilde{{\bm \psi}} + \tilde{{\bm n}} = 0 ,
\end{equation}
where the drift and noise terms are given by
\begin{equation}
  \label{eq:D0n}
  {\bm D}  = {\bm D}_0 \equiv
  \begin{pmatrix}
    0 & ik\\
   ik\vs^2 & 1-\vs^2
  \end{pmatrix}
,\qquad \tilde{{\bm n}} =
\begin{pmatrix}
  1\\ ik 
\end{pmatrix}
\tilde{f} \,.
\end{equation}
The solution to these equations for arbitrary noise and drift, with the given initial conditions, can be written as
\begin{equation}
  \label{eq:q-sol}
  \tilde{{\bm \psi}}(k,\tau) = - \int_{\tau_0}^\tau \frac{d\tau'}{\tau'} \tilde{{\bm U}}(k;\tau,\tau') \tilde{{\bm n}}(k,\tau')
\end{equation}
where $\tilde{{\bm U}}$ is the evolution operator satisfying
\be
\tau \frac{\partial \tilde{{\bm U}}(k;\tau,\tau')}{\partial \tau} + {\bm D}(k,\tau) \tilde{{\bm U}}(k;\tau,\tau')=0
\ee
subject to the condition $\tilde{{\bm U}}(k;\tau,\tau)=1$.  Explicitly
\begin{equation}
\label{eq:sol-U}
\tilde{{\bm U}}(k;\tau,\tau') = {\cal T}\exp\left\{-\int_{\tau'}^{\tau} \frac{d\tau''}{\tau''} {\bm D}(k,\tau'') 
\right\}
\end{equation}
where ${\cal T}$ is the time ordering operator.  Comparing  Eq. (\ref{eq:q-sol}) to  Eqs. (\ref{eq:rho-G-f}) and (\ref{eq:omega-G-f}) allows for
determination of $\tilde{G}_{\rho}$ and $\tilde{G}_{\omega}$, namely
\begin{equation}
 \begin{pmatrix}
      \tilde{G}_{\rho}(k;\tau,\tau')\\
      \tilde{G}_{\omega}(k;\tau,\tau')
  \end{pmatrix}
 = \tilde{\bm U}(k;\tau,\tau')
 \begin{pmatrix}
  1\\ik
 \end{pmatrix}.
\label{eq:G-rho-U}
\end{equation}

The second method is to eliminate one of the variables in favor of the other to arrive at a single second order differential equation.  Elimination of $\tilde{\omega}$ results in
\be
\tau^2 \frac{\partial^2 \tilde{\rho}}{\partial \tau^2} + (2-{\vs}^2) \tau \frac{\partial \tilde{\rho}}{\partial \tau} + {\vs}^2 k^2 \tilde{\rho} + \left[ \tau \frac{\partial \tilde{f}}{\partial \tau} + \left( k^2 + 1-{\vs}^2 \right) \tilde{f} \right] = 0 \,.
\label{2ndorder}
\ee
Denote the two independent solutions to the homogeneous equation, when $\tilde{f}=0$, by $\tilde{\rho}_1$ and $\tilde{\rho}_2$.  The function $\tilde{G}_{\rho}$ is constructed from a linear combination of the two homogeneous solutions as
\be
\tilde{G}_{\rho}(\tau,\tau') = \tilde{a}_1(\tau') \tilde{\rho}_1(\tau) + \tilde{a}_2(\tau') \tilde{\rho}_2(\tau) \, .
\label{G2ndorder}
\ee
The functions $\tilde{a}_1$ and $\tilde{a}_2$ are determined by substitution into Eq. (\ref{2ndorder}), with the result that
\be
\tilde{a}_1(\tau) \tilde{\rho}_1(\tau) + \tilde{a}_2(\tau) \tilde{\rho}_2(\tau) = 1
\label{G2ndordersolve1}
\ee
\be
\tilde{a}_1(\tau) \tau \frac{\partial \tilde{\rho}_1(\tau)}{\partial \tau} +
\tilde{a}_2(\tau) \tau \frac{\partial \tilde{\rho}_2(\tau)}{\partial \tau} = k^2
\label{G2ndordersolve2}
\ee 
Solution of this pair of algebraic equations solves the problem of determining $\tilde{G}_{\rho}$.  The Green function for $\omega$ is then found by substitution into Eq. (\ref{11eqF}) to yield
\be
\tilde{G}_{\omega}(k;\tau,\tau') = \frac{i \tau}{k} \frac{\partial \tilde{G}_{\rho}(k;\tau,\tau')}{\partial \tau} \, .
\ee
Without an explicit equation of state and viscosities it is not possible to be more specific.

\subsection{Linear equation of state}

To proceed further we now choose the equation of state $P = {\vs}^2 \epsilon$ with a constant speed of sound ${\vs}$.  This is a reasonable approximation to QCD at high temperature.  The response functions $\tilde{G}_{\rho}$ and $\tilde{G}_{\omega}$ can then be found in terms of elementary functions. 

In the Langevin approach the drift matrix ${\bm D}$ is a constant, and
the evolution operator $\tilde{{\bm U}}$ can be found explicitly in
terms of the eigenvalues $\lambda_{\pm}$ and eigenvectors ${\bm \psi}_{{\pm}}$ of ${\bm D}$.  With
\begin{equation}
\label{eq:D-ev}
{\bm D} \tilde{{\bm \psi}}_{{\pm}} = \lambda_{\pm} \tilde{{\bm \psi}}_{{\pm}}
\end{equation}
we find (cf. Ref.\cite{Baym:1984sr})
\begin{align}
\label{eq:lambda12}
\lambda_{\pm} =& \alpha \pm \beta \nonumber \\
\alpha =& \frac12\left( 1-\vs^2 \right) \nonumber \\
\beta =&  \sqrt{\alpha^2-\vs^2k^2}
\end{align}
and
\begin{equation}\label{eq:psi-pm}
\quad  \tilde{{\bm \psi}}_{\pm} =  (ik,\lambda_{\pm}) \,.
\end{equation}
Then the evolution matrix $\tilde{{\bm U}}$ can be determined as
\be
\label{eq:U-matrix}
\tilde{{\bm U}}(k;\tau,\tau') = \frac{(\tau'/\tau)^{\lambda_-}}{\lambda_+-\lambda_-}
  \begin{pmatrix}
    \lambda_+ & 
-ik \\ 
-ik\vs^2& 
- \lambda_-\\
  \end{pmatrix}
-  \frac{(\tau'/\tau)^{\lambda_+}}{\lambda_+ - \lambda_-}
  \begin{pmatrix}
    \lambda_- & 
-ik \\ 
-ik\vs^2& 
- \lambda_+\\
  \end{pmatrix} \, .
\ee
The response functions are then given by Eqs. (\ref{eq:G-rho-U}) as
\begin{equation}
\tilde{G}_{\rho}(k;\tau,\tau') =  \left(\frac{\tau'}{\tau}\right)^{\alpha} \left[ \cosh\left(\beta \ln(\tau/\tau')\right) +
\left( \frac{\alpha +k^2}{\beta} \right) \sinh\left(\beta \ln(\tau/\tau')\right)  \right]
\label{eq:G-rho}
\end{equation}
\begin{equation}
\tilde{G}_{\omega}(k;\tau,\tau') = ik \left(\frac{\tau'}{\tau}\right)^{\alpha}  \left[ \cosh\left(\beta \ln(\tau/\tau')\right) -
\left( \frac{\alpha +{\vs}^2}{\beta} \right) \sinh\left(\beta \ln(\tau/\tau')\right)  \right]
\label{eq:G-omega}
\end{equation}
Note that $\beta$ is real if $|k| < (1-{\vs}^2)/2{\vs}$ and is pure imaginary if $|k| > (1-{\vs}^2)/2{\vs}$ leading to exponential or oscillatory behavior, respectively.

In the second method, the two solutions to Eq. (\ref{2ndorder}) are found to be
\begin{align}
\tilde{\rho}_1(\tau) =& \left(\frac{\tau_0}{\tau}\right)^{\alpha + \beta} \nonumber \\
\tilde{\rho}_2(\tau) =& \left(\frac{\tau_0}{\tau}\right)^{\alpha - \beta} \, .
\label{rhosolutioninviscid}
\end{align}
The corresponding coefficient functions to construct $\tilde{G}$ are
\ba
\tilde{a}_1(\tau') &=& \frac{\beta - \alpha -k^2}{2\beta}
\left(\frac{\tau'}{\tau_0}\right)^{\alpha + \beta} \nonumber \\
\tilde{a}_2(\tau') &=& \frac{\beta + \alpha +k^2}{2\beta}
\left(\frac{\tau'}{\tau_0}\right)^{\alpha - \beta} \, .
\ea
The response functions are identical to the ones given in Eqs. (\ref{eq:G-rho}) and (\ref{eq:G-omega}).

\subsection{Singularities and the sound horizon}
\label{sec:sing-sound-horiz}

It is instructive to analyze the singularities of $G_{\rho}(\xi;\tau,\tau')$ and $G_{\omega}(\xi;\tau,\tau')$.  As we shall see, they reflect propagation of sound along the $z$ axis on top of the expanding medium. The propagation of sound waves in the transverse directions has been discussed in Ref. \cite{Staig:2010pn}.

First, observe that $\tilde{G}_{\rho}$ and $\tilde{G}_{\omega}$ are meromorphic functions of $k$.  The sole singularity is an essential singularity at infinity.  As $k \to \infty$, $\beta \to i\vs k$ whereas $\alpha$ remains a constant.  Therefore, when $|\xi|>\vs \log(\tau/\tau')$, one can close the contour in the Fourier integral over $k$ around either the upper or lower large semi-circle (depending on the sign of $\xi$) to show that
\begin{equation}
  \label{eq:horizon}
   {G}_X(\xi;\tau,\tau')=0 \quad\mbox{when}\quad |\xi|>\vs \ln(\tau/\tau') \, .
\end{equation}
This means that there is a sound horizon which expands logarithmically with proper time~$\tau$. This result is confirmed by the observation that, in the local rest frame the velocity of the front, $\tau d\xi/d\tau$, equals $\vs$.
Arguing similarly, or using Eq. (\ref{eq:G-GG-xi}) directly, one can show that the correlations do not spread beyond the sound horizon:
\begin{equation}
  \label{eq:horizon-XY}
   {G}_{XY}(\xi;\tau,\tau')=0 \quad\mbox{when}\quad |\xi|>2\vs \ln(\tau/\tau') \, .
\end{equation}

The singularities of ${G}_{XY}(\xi;\tau,\tau')$ at the sound horizon can be analyzed by considering the large $k$
asymptotics of its Fourier transform. For example, since for large $k$
\begin{equation}\label{eq:G_rho-large-k}
\tilde{G}_{\rho}(k;\tau,\tau') \rightarrow \frac{k}{{\vs}} \left( \frac{\tau'}{\tau} \right)^{\alpha} \sin\left[ {\vs} k \ln(\tau/\tau') \right] \, ,
\end{equation}
we can use Eq. (\ref{eq:G-GG}) to find that
\bd
G_{\rho\rho}(\xi;\tauf,\tau) =
\int_{-\infty}^{\infty} \frac{dk}{2\pi} {\rm e}^{ik\xi} \tilde{G}_{\rho}(k;\tauf,\tau) \tilde{G}_{\rho}(-k;{\tauf},\tau) \rightarrow
\ed
\be
\frac{1}{4\vs^2} \left( \frac{\tau}{\tauf} \right)^{2\alpha}
\left[ \delta^{\prime\prime}(\xi - 2\vs \ln(\tauf/\tau)) +  \delta^{\prime\prime}(\xi + 2\vs \ln(\tauf/\tau)) - 2  \delta^{\prime\prime}(\xi)  \right] 
\label{eq:G-rhorho-sing}
\ee
where $\xi = \xi_2 - \xi_1$.  We see that the correlator is singular whenever there is a noise source event at earlier time $\tau$ such that a sound cone originating from it goes through both points $\xi_1$ and $\xi_2$ at time $\tauf$.  That source point is located midway between the points $\xi_1$ and $\xi_2$ for the first two 
delta-functions in Eq. (\ref{eq:G-rhorho-sing}). For the last delta-function, there are two source events located a distance $\ln(\tauf/\tau)$ away on either side of the coinciding points $\xi_1=\xi_2$.

The second derivative of the delta-function can be traced back to the derivative of the force term $f$ with respect to $\xi$.  This derivative is a consequence of momentum conservation.  The second derivatives of the delta-function in Eq. (\ref{eq:G-rhorho-sing}) only represent the leading
singularities. There exist sub-leading singularities, such as step-functions, delta-functions and first derivatives of delta-functions.  The prefactor $(\tau/\tauf)^{2\alpha}$ describes the dilution of the fluctuation due to the expansion.

\subsection{The wake}
\label{sec:wake}

In a stationary medium, and neglecting viscous effects, sound propagation would be the only source of correlations. This is because
the sound propagation would be non-dispersive with a linear relation between frequency and wavenumber.  In the expanding medium we are
considering, sound propagation also leads to a sound horizon but, unlike the stationary case, the dispersion relation is non-linear
according to eq. (\ref{eq:lambda12}). This leads to a wake behind the sound front, even without dissipation.

\begin{figure}
  \centering
  \includegraphics{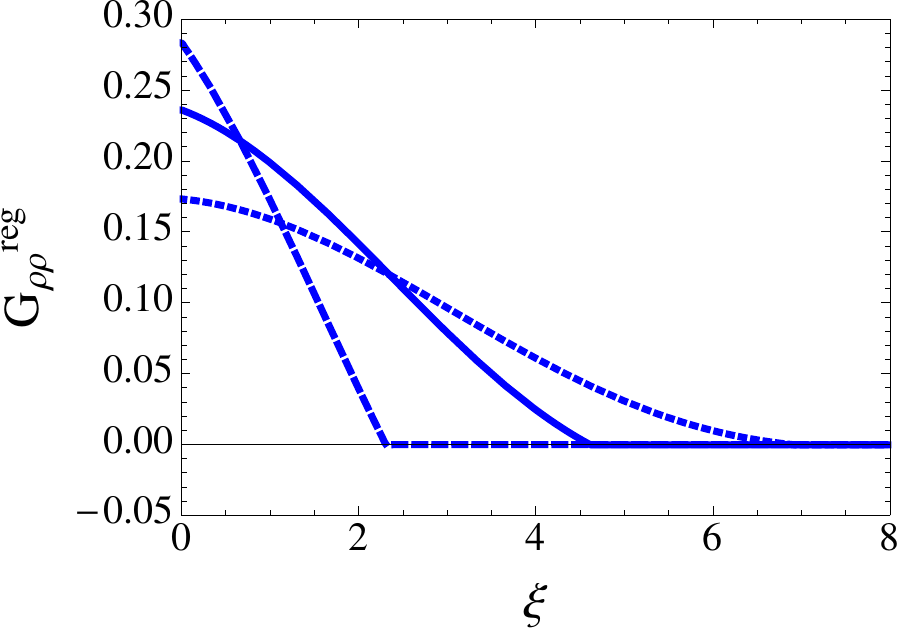}
  \caption{The regular (continuous) part of the correlator $G_{\rho\rho}(\xi;\tauf,\tau)$ with ${\vs}^2=1/3$.
  The correlator is shown for $\ln(\tauf/\tau)=2$ (dashed line),  $\ln(\tauf/\tau)=4$ (solid line),
  and  $\ln(\tauf/\tau)=6$ (dotted line).}
  \label{fig:G_reg}
\end{figure}

Let us illustrate this using Eq. (\ref{eq:G-rho}) to calculate the correlator
$\tilde G_{\rho\rho}(k;\tauf,\tau)=|\tilde G_{\rho}(k;\tauf,\tau)|^2$.  
The Fourier transform $G_{\rho\rho}(\xi;\tauf,\tau)$ of this function is singular as
discussed in the previous section.  It is instructive to separate the most singular part of this function
by expanding $G_{\rho\rho}(k;\tauf,\tau)$ in powers of $1/k^2$ and keeping all terms regular at $k=0$.  This leads to
\ba
  \label{eq:Gsing-k}
  {\tilde  G}_{\rho\rho}^{\rm sing}(k;\tauf,\tau)
&=& \left(a_1k^2+b_1\right) + \left(a_2k^2+b_2\right)\cos\left(2\vs \ln(\tauf/\tau) k\right) 
\nonumber \\
&& + \left(\frac{a_3k^2+b_3}{k}\right)\sin\left(2\vs \ln(\tauf/\tau) k\right) ,
\ea
where the expansion coefficients $a_n$ and $b_n$ are functions of $\ln(\tauf/\tau)$.
The Fourier transform of Eq. (\ref{eq:Gsing-k}),  $G_{\rho\rho}^{\rm sing}(\xi;\tauf,\tau)$, is a sum of step functions and its
derivatives with singularities located at $\xi=0$ and $\xi = \pm2\vs \ln(\tauf/\tau)$. The
most singular term we have already written out in Eq. (\ref{eq:G-rhorho-sing}).
The regular part
\begin{equation}
  \label{eq:G-reg}
G_{\rho\rho}^{\rm reg}\equiv{  G}_{\rho\rho}-G_{\rho\rho}^{\rm sing}
\end{equation}
is a continuous function of $\xi$, which is shown in Fig. \ref{fig:G_reg}.
We see that the correlations spread along the rapidity axis from $\xi=0$ with time. In Appendix \ref{sec:long-time-limit} we show that for long times this process resembles diffusion, and for asymptotically large $\tauf/\tau$ the function $G_{\rho\rho}$ is given by a Gaussian. However, one should bear in mind that this diffusion is occurring in the absence of any dissipative effects since we have neglected viscous terms in the hydrodynamic equations at this stage.

\begin{figure}
  \centering
  \includegraphics{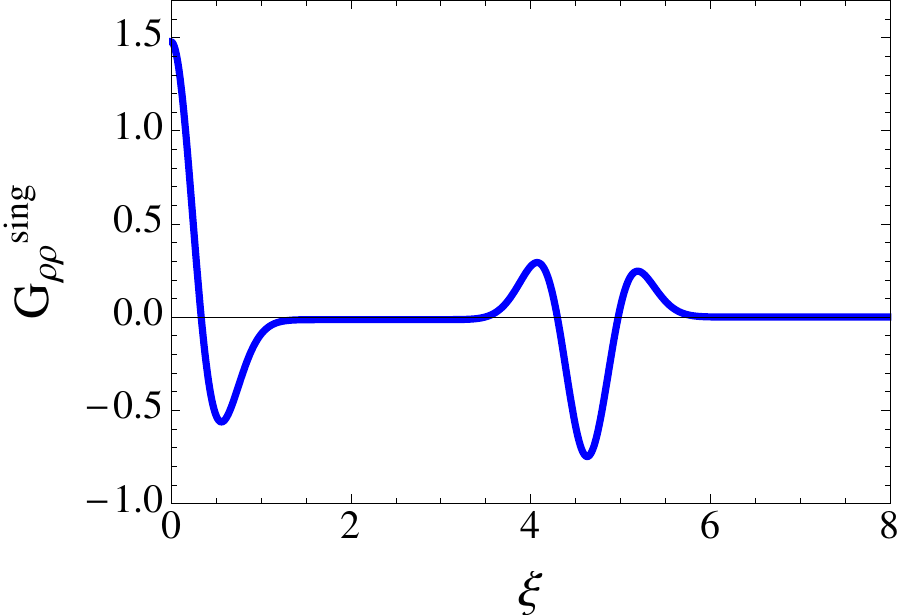}
  \caption{The singular part of the correlator $G_{\rho\rho}(\xi;{\tauf},\tau)$ with ${\vs}^2=1/3$ and $\ln({\tauf}/\tau)=4$ . 
    The function is smeared by a Gaussian of variance $\sigma^2=0.1$ in order to
    show the nature of the singularities.}
  \label{fig:G_sing}
\end{figure}

In order to display the singular part $G^{\rm sing}_{\rho\rho}$ we convolute it with the Gaussian
\bd
\frac{1}{\sqrt{2\pi}\sigma} {\rm e}^{-(\xi-\xi')^2/2\sigma^2}
\ed
of a small width $\sigma^2=0.1$. This simply replaces
  delta-functions with Gaussians and step functions with  error
  functions.  The result is shown in Fig. \ref{fig:G_sing}.

It is useful to note that the correlator $G_{\rho\rho}(\xi;{\tauf},\tau)$ obeys the sum rule
\begin{equation}
  \label{eq:sr}
  \int_{-\infty}^{\infty} d\xi\,G_{\rho\rho}(\xi;\tauf,\tau)=1 \,,
\end{equation}
which is related to entropy conservation. Indeed, the time-independence of the integral in Eq. (\ref{eq:sr}) is due to the zero mode in Eq. (\ref{eq:D-ev}):
$\lambda_-(k=0)=0$. The origin of the zero mode can be understood using the equation of motion (\ref{eq:dts/dt}).   Expressing the fluctuation at $k=0$ as $\delta\tilde \rho(k=0,\tau)={\rm const}\int \delta s(\xi,\tau) \,\tau d\xi$, we see that it is proportional to the fluctuation of the total entropy whose relaxation rate must vanish in the inviscid case. It is also interesting to note that at asymptotically large times $\tauf/\tau\gg1$ the sum rule is saturated by the regular part of $G_{\rho\rho}$ since the singular part is suppressed by a factor $(\tau/\tauf)^{2\alpha}$; see Eq. (\ref{eq:G-rhorho-sing}) and Appendix \ref{sec:long-time-limit}.

\subsection{Viscosity and taming of the singularities}
\label{sec:viscosity}

We now include the effects of viscosity on the space-time evolution of the system.  In general, viscosity acts to smooth out gradients in temperature and flow velocity.  Even if the viscosity is very small and its effects on the solutions to the equations of motion can mainly be neglected, the effect of viscosity on the correlation functions
will still be important near the sound horizon singularities discussed in the previous two sections. The goal of this section is to demonstrate how the effect of viscosity smoothes out these singularities.  The equations to be solved now are (\ref{11eq}) and (\ref{22eq}).  For simplicity, and for illustrative purposes, we consider a constant value of $\nu$ as well as a linear equation of state, $\vs^2=$ constant. 
 
As in the inviscid case, we combine Eqs. (\ref{11eq}) and (\ref{22eq})
into a two-component matrix Langevin equation. It is convenient to use a rescaled variable $\omega\,(1-\nu/T\tau)$ instead of $\omega$. The drift operator now has the form 
${\bm D} = {\bm D}_0 + {\bm D}_1$ with ${\bm D}_0$ being given by
Eq. (\ref{eq:D0n}) and the viscous contribution (neglecting terms
higher order in $\nu/T\tau$)
\begin{equation}
  \label{eq:Dn}
  \bm D_1 =\frac{\nu}{T\tau}
  \begin{pmatrix}
    \vs^2 & -ik\\
    -ik & 1+\vs^2+k^2
  \end{pmatrix} \, .
\end{equation}
To see the effect of viscosity more clearly it is convenient to rewrite the evolution operator in Eq. (\ref{eq:sol-U}) as
\begin{equation}
  \label{eq:U-DI}
    \tilde{\bm U}(k;\tau,\tau') = (\tau_0/\tau)^{\bm D_0}\,
{\cal T} \exp\left\{ -\int_{\tau'}^{\tau} \frac{d\tau''}{\tau''} \,\bm D_I(\tau'') \right\}\,
(\tau'/\tau_0)^{\bm D_0} ,
\end{equation}
where we defined the matrix $\bm D_I$ as
 \begin{equation}
  \label{eq:DI}
  \bm D_I(\tau'') = (\tau''/\tau_0)^{\bm D_0} \bm D_1(\tau'')\, (\tau_0/\tau'')^{\bm D_0}.
\end{equation}
Note that $\bm D_I= {\cal O}(\nu)$. To leading order in $\nu$ (in the exponent) we can use the Baker-Campbell-Hausdorff formula to remove the operation of 
time ordering in Eq.~(\ref{eq:U-DI}) which greatly simplifies calculations. The matrix $\tilde{\bm U}$ can be then calculated through a lengthy matrix algebra.

In order to evaluate the effect of dissipation on the delta-function singularities in $\xi$-space let us consider the limit $k\to\infty$. In this regime the matrix algebra 
simplifies and can be performed more easily.  Keeping only the leading terms\footnote{More precisely, the limit we consider is $k\gg1$ but $\nu k^2\ll T\tau$.} in Eq. (\ref{eq:Dn}) we obtain:
\begin{equation}
  \label{eq:D-large-k}
  \bm D_0 \approx
\begin{pmatrix}
    0 & ik\\
    ik\vs^2 &0
  \end{pmatrix}
,\quad
  \bm D_1 \approx \frac{\nu}{T\tau}
  \begin{pmatrix}
    0 & 0\\
    0 & k^2
  \end{pmatrix} 
\equiv  \frac{\nu\, k^2}{2\, T\tau} ({\bm 1} - {\bm\sigma}_3) ,
\end{equation}
where ${\bm\sigma}_3$ denotes the third Pauli matrix. We next note that ${\bm\sigma}_3$
anticommutes with ${\bm D}_0$ in Eq.~(\ref{eq:D-large-k}), which implies that Eq. (\ref{eq:DI}) can be written as 
 \begin{equation}
  \label{eq:DI-large-k}
  \bm D_I(\tau'') \approx \frac{\nu\, k^2}{2T(\tau'')\tau''}\,
  \left[ {\bm 1} - {\bm\sigma}_3\, (\tau_0/\tau'')^{2\bm D_0} \right] .
\end{equation}
Since the eigenvalues of ${\bm D}_0$ are pure imaginary ($\pm ik{\vs}$), the second term in brackets in Eq. (\ref{eq:DI-large-k}) is an oscillating function of
$\ln\tau''$.  Upon integration over $\tau''$ in Eq. (\ref{eq:U-DI}) the oscillating terms will be suppressed by a power of $1/k$, 
and the leading contribution, of order $k^2$, will be proportional to the unit matrix, an important simplification.

The integral needed then is
\begin{equation}
  \label{eq:HT1T2}
H(\tau,\tau') \equiv \int_{\tau'}^{\tau} \frac{d\tau''}{\tau''}\frac{1}{T(\tau'')\tau''} =
 \frac{1}{2\alpha} \left[ \frac{1}{T(\tau')\tau'} - \frac{1}{T(\tau) \tau} \right]
\end{equation}
where we used $T=T_0 (\tau_0/\tau)^{\vs^2}$ and $\alpha$ defined in Eq. (\ref{eq:lambda12}). Thus
we can write
\begin{equation}
  \label{eq:U2}
  {\tilde{\bm U}}(k;\tau,\tau') \approx (\tau_0/\tau)^{\bm D_0}\, {\rm e}^{ - \nu H(\tau,\tau')k^2/2}\,  
  (\tau'/\tau_0)^{\bm D_0} 
= {\rm e}^{ - \nu H(\tau,\tau')k^2/2}\, \tilde{\bm U}_0(k;\tau,\tau')
\end{equation}
where $\tilde{\bm U}_0(k;\tau,\tau')=(\tau'/\tau)^{\bm D_0}$ is the evolution matrix in the inviscid case.  We see that the effect of viscosity is to dampen the oscillatory (or constant) behavior at large $k$ as long as $\tau\neq\tau'$.  The overall result is that both $\tilde{G}_{\rho}(k;\tau,\tau')$ and
$\tilde{G}_{\omega}(k;\tau,\tau')$ are to be multiplied by the Gaussian function ${\rm e}^{ - \nu H(\tau,\tau')k^2/2}$.  The effect of viscosity is thus simply a diffusion in space-time rapidity whose proper-time dependence is controlled by the function $H(\tau,\tau')$.

Next we turn to the second method for solving the same problem.  In the limit of small viscosity, $\nu \ll \tau T$, Eqs. (\ref{11eq}-\ref{22eq}) in $k$-space become
\be
\label{eq:rho-eq}
\tau \frac{\partial \tilde{\rho}}{\partial \tau} + ik\tilde{\omega} + \tilde{f} = 0
\ee
\be
\tau \frac{\partial \tilde{\omega}}{\partial \tau} + (1-{\vs}^2)\tilde{\omega} + ik\left( {\vs}^2\tilde{\rho} + \tilde{f} \right) + \frac{\nu k^2}{\tau T} \tilde{\omega}= 0 \,.
\ee
In the limit $\nu=0$ these reduce to exactly the same equations as studied earlier.  The only new term is the last one in the second equation above on account of the fact that even though $\nu/\tau T$ is assumed to be small, large enough values of $k$ will make it important.  Upon eliminating $\tilde{\omega}$ one arrives at a single second order differential equation.
\be
\tau^2 \frac{\partial^2 \tilde{\rho}}{\partial \tau^2} + \left( 2-{\vs}^2 +\frac{\nu k^2}{\tau T} \right) \tau \frac{\partial \tilde{\rho}}{\partial \tau} + {\vs}^2 k^2 \tilde{\rho} + \left[ \tau \frac{\partial \tilde{f}}{\partial \tau} + \left( k^2 + 1-{\vs}^2 \right) \tilde{f} \right] = 0 \,.
\label{2ndorderviscous}
\ee
Compared to Eq. (\ref{2ndorder}) there is only one new term.

To find the solutions to the homogeneous equation ($\tilde{f}=0$) it is convenient to change variables from $\tau$ to $x = (\tau_0/\tau)^{2\alpha}$.  This leads to the second order differential equation 
\be
\frac{\partial^2 \tilde{\rho}}{\partial x^2} - \frac{\nu k^2}{2\alpha \tau_0 T_0} \frac{\partial \tilde{\rho}}{\partial x}+  \frac{{\vs}^2 k^2}{4\alpha^2 x^2} \tilde{\rho} = 0 \, .
\label{eq:deltas-tilde}
\ee 
The two independent solutions to this equation are
\begin{align}
  \label{eq:rho-solutions}
  \tilde\rho_1\sim& \sqrt x 
\exp\left(\frac{\nu k^2 x}{4\alpha \tau_0 T_0} \right)
K_{\beta/2\alpha}\left( \frac{\nu k^2 x}{4\alpha \tau_0 T_0}
\right) \, , \;\\
  \tilde\rho_2\sim& \sqrt x 
\exp\left(\frac{\nu k^2 x}{4\alpha \tau_0 T_0} \right)
I_{\beta/2\alpha}\left( \frac{\nu k^2 x}{4\alpha \tau_0 T_0}
\right) \, . \;
\end{align}
For $\nu k^2/\tau T \ll 1$, the arguments of both the Bessel functions and the exponential are small.  Keeping the lowest order terms in both functions we obtain the same result in the inviscid case as given by Eq. (\ref{rhosolutioninviscid}).  When $k$ is sufficiently large, $\nu k^2/\tau T$ may not be small. In this regime, however, the index of the Bessel functions becomes large, $\beta/2\alpha \sim{\cal O}(k)$, and the Bessel function can still be approximated\footnote{This requires only that $z^2\ll\mu$, i.e. the argument of the Bessel function does not have to be small if the index is large.} as $K_\mu(z)\sim z^{-\mu}$, $I_\mu(z)\sim z^\mu$. 
Therefore the solutions are approximately (the normalization doesn't matter)
\ba
\tilde{\rho}_1(\tau) &=& \left(\frac{\tau_0}{\tau}\right)^{\alpha +
  \beta} \exp\left(\frac{\nu k^2}{4\alpha \tau T(\tau)} \right)  \nonumber \\
\tilde{\rho}_2(\tau) &=& \left(\frac{\tau_0}{\tau}\right)^{\alpha - \beta} \exp\left(\frac{\nu k^2}{4\alpha\tau T(\tau)} \right) \, .
\label{rhosolutionviscous}
\ea
These are the same as the inviscid case Eqs. (\ref{rhosolutioninviscid}) with an additional exponential factor. 

The Green function $\tilde{G}_{\rho}$ is constructed in exactly the same way as in the inviscid case, Eqs. (\ref{G2ndorder}-\ref{G2ndordersolve2}) because the homogeneous term in $\tilde{f}$ is unchanged.  This now leads to
\ba
\tilde{a}_1(\tau') &=& \frac{1}{2\beta} \left[ \beta -\alpha -k^2 -\frac{\nu k^2}{2\tau' T(\tau')} \right]
\left(\frac{\tau'}{\tau_0}\right)^{\alpha + \beta}
\exp\left(-\frac{\nu k^2}{4\alpha\tau' T(\tau')} \right) \nonumber \\
\tilde{a}_2(\tau') &=& \frac{1}{2\beta} \left[ \beta +\alpha +k^2 +\frac{\nu k^2}{2\tau' T(\tau')} \right]
\left(\frac{\tau'}{\tau_0}\right)^{\alpha - \beta}\exp\left(-\frac{\nu k^2}{4\alpha\tau' T(\tau')} \right) \, .
\ea
The result for $\tilde{G}_{\rho}$ is
\bd
\tilde{G}_{\rho}(k;\tau,\tau') =  \left(\frac{\tau'}{\tau}\right)^{\alpha} \left[ \cosh\left(\beta \ln(\tau/\tau')\right) + \frac{1}{\beta}
\left( \alpha +k^2 + \frac{\nu k^2}{2\tau' T(\tau')} \right) \sinh\left(\beta \ln(\tau/\tau')\right)  \right] 
\ed
\be
\times \exp\left[-\frac{\nu k^2}{4\alpha} \left( \frac{1}{\tau' T(\tau')} -  \frac{1}{\tau T(\tau)}\right) \right]
\label{eq:G-rho-nu}
\ee
When $\nu = 0$ it reduces to the inviscid case represented by Eq. (\ref{eq:G-rho}).  In the regime  $\nu k^2/\tau T \ll 1$ the term proportional to $\nu$ in front of
the $\sinh$ is negligible and this result coincides with Eqs. (\ref{eq:HT1T2}, \ref{eq:U2}). The Eq. (\ref{eq:G-rho-nu}) is, however, somewhat more accurate, since
it does not assume $\nu k^2/\tau T\ll1$, which is reflected in the coefficient of the $\sinh$.

\subsection{Example of other sources of smoothing of singularities}
\label{sec:noise-broadening}
 
The sound horizon singularities would be smeared if the delta-function correlator for the noise in Eq. (\ref{eq:ff-xi}) were replaced by a narrowly peaked regular
function. Indeed, the origin of the noise is the fluctuation on a microscopic scale whose correlation length, small on the hydrodynamic scale, is non-zero. Here we shall consider the effect of the finite correlation length of the noise.  Although the introduction of a finite correlation length is physically intuitive, it should be borne in mind that from the point of view of hydrodynamics it corresponds to inclusion of some, but not all, higher-order corrections in the systematic gradient expansion. We shall use this effect in the next section to estimate possible sensitivity of our results to higher-order hydrodynamic corrections.

The most microscopically sensible replacement for the delta-functions is one that is exponential in time and in space. When distances are small the time and space intervals can be expressed in terms of Bjorken coordinates as $\Delta t\approx\Delta\tau$ and $\Delta z\approx\tau\Delta\xi$.  For simplicity,  we keep the delta-function
$\delta(\tau_2-\tau_1)$ in Eq. (\ref{eq:ff-xi}) but replace $\delta(\xi_2-\xi_1)/\tau_1$ with an exponential
\be
\frac{\delta(\Delta \xi)}{\tau} \rightarrow \frac{{\rm e}^{-\tau |\Delta \xi|/\lambda(\tau)}}{2\lambda(\tau)} \, ,
\ee
where $\lambda(\tau)$ denotes the correlation length, which is a function of proper time.
The net result is to multiply the right-hand side of Eq. (\ref{k-correlation}) by
\bd
\frac{\tau_1^2}{\tau_1^2 + \lambda^2(\tau_1) k_1^2} \,.
\ed
It is natural to assume that $\lambda(\tau) = c_{\lambda}/T(\tau)$.  We can make a simple estimate of the constant $c_{\lambda}$ by assuming that $\lambda$ is given by the average interparticle distance at temperature $T$.  For gluons plus three flavors of massless quarks the particle density is
\be
n = \frac{127 \zeta(3)}{4\pi^2} T^3
\ee
which gives $c_{\lambda}=n^{-1/3}T = 0.637$.  The implications of this choice will be investigated in the next section.

\section{Phenomenology}
\label{sec:pheno}

In order to make contact with experiment we need to consider how the fluctuations are frozen out.  In general, hydrodynamic freeze-out
occurs when the particles can no longer maintain local thermal equilibrium and therefore begin free-streaming.  A schematic approach to the freeze-out problem is represented by the Cooper-Frye formula \cite{Cooper:1974mv} which describes the distribution of emitted particles as an integral over a freeze-out hypersurface $\Sigma_{\rm f}$, usually chosen to coincide with a surface of constant temperature $\Tf$ (isothermal freeze-out) of the expanding fluid.  In the Bjorken
expansion scenario this surface is also a $\tauf=$ constant surface, but this equivalence holds only for averaged quantities. The fluctuations of temperature mean that the conditions $\Tf=$ constant and $\tauf=$ constant differ. In this paper we shall choose the simplest of these two conditions, isochronous freeze-out with $\tauf=$ constant, which has been used for the study of fluctuations in \cite{Asakawa:2010bu}.  The alternative approach was pursued by Staig and Shuryak in their treatment of initial state fluctuations in the transverse space \cite{Staig:2011wj}. Both approaches offer only a schematic representation of freeze-out, but sufficient for our illustrative purposes. We leave the proper implementation of freeze-out to future studies.

\subsection{Freeze-out and rapidity smearing}
\label{sec:freeze}

We start with the Cooper-Frye formula \cite{Cooper:1974mv} for the phase space distribution of emitted particles of a given species
\be
\label{eq:dNd3p}
p^0\frac{dN_{\rm s}}{d^3p} 
= \int_{\Sigma_{\rm f}} d^3\sigma_\mu\, p^\mu\,  
\theta(\sigma \cdot p)  \, \ds \, f_{\rm s}({\bm x},{\bm p}),
\ee
where 
\be
\label{eq:fp}
f_{\rm s}({\bf x},{\bf p}) = \left( {\rm e}^{(p\cdot u - \mu_{\rm s})/T} \pm 1 \right)^{-1}
\ee
denotes the local thermal distribution for the particle species with degeneracy $\ds$, and the step function ensures emission in the forward or outward direction. 
For the isochronous freeze-out in proper time the freeze-out hypersurface is given by $\tau=\tauf=$ constant. This means that the four-vector $d^3\sigma_\mu$ has only a $\tau$-component in the Bjorken coordinates. Thus $d^3\sigma\cdot p = p^\tau \tau d\xi d^2x_\perp$, where the $\tau$-component of $p^\mu$ is $p^\tau=m_\perp \cosh(\eta-\xi)$ with $m_\perp=\sqrt{p_\perp^2+m_0^2}$. Furthermore, $\theta(\sigma_\mu p^\mu) = \theta({p^\tau}) = 1$.
For simplicity, we set $\mu_{\rm s}=0$ and neglect the quantum correction $\pm 1$ in the phase space distribution (\ref{eq:fp}).  For the purpose of obtaining numerical values we consider charged pions, $\ds=2$.  We are interested in the number of particles of a given species per unit {\em kinematic} rapidity $\eta = \tanh^{-1}(p_z/p_0)$. Integrating (\ref{eq:dNd3p}) over transverse area $\aperp$ and using $(u^\tau,u^\xi)=(\cosh\omega,\tau^{-1}\sinh\omega)$,
  $(p^\tau,p^\xi)=m_\perp(\cosh(\eta-\xi),\tau^{-1}\sinh(\eta-\xi))$, we obtain
\begin{equation}
  \label{eq:dNdeta-A}
  \frac{dN}{d\eta} = \ds \int \frac{d^2 p_\perp \aperp}{(2\pi)^3}\,
  \tauf m_\perp \int d\xi \cosh(\eta-\xi) \, 
  \exp\left[-m_\perp \cosh(\eta-\xi-\omega)/T\right] \,.
\end{equation}

Fluctuations in $dN/d\eta$ are caused by fluctuations in temperature $T$ (around $\Tf$) and flow rapidity $\omega$.  Expanding to linear order in fluctuations and integrating over $d^2p_\perp$ we find
\begin{equation}
  \label{eq:dNdeta-tilde}
    \delta\left(\frac{d N}{d\eta}\right) =  \frac{\ds \aperp\tauf \Tf^3}{(2\pi)^2}\,
\int d\xi\,
\,\frac{ \rho\, \vs^2 + \omega\, \tanh(\eta-\xi)}{\cosh^2(\eta-\xi)}
\,\Gamma\left(4,\frac{m_0}{\Tf}\cosh(\eta-\xi)\right)
\end{equation}
where $\Gamma(x,y)$ is incomplete Gamma function and we used
\begin{equation}
  \label{eq:1}
 \delta T/T = \vs^2 \delta s/s= \vs^2\rho \,.
\end{equation}
The rapidity correlator can then be written as
\begin{equation}
  \label{eq:dNdeta-dNdeta}
  \la \delta\frac{d N}{d\eta_1}\,
\delta\frac{d N}{d\eta_2}\ra =
\left(\frac{\ds \aperp\tauf \Tf^3}{(2\pi)^2}\right)^2\,
\int d\xi_1 \int d\xi_2 \sum_{\substack{X=\rho,\omega\\Y=\rho,\omega}}
F_X(\eta_1-\xi_1)F_Y(\eta_2-\xi_2)\,\Corr_{XY}(\xi_1-\xi_2;\tauf)
\end{equation}
where the smearing functions are given by
\begin{eqnarray}
  \label{eq:F-rho}
 F_\rho(x) = \frac{\vs^2}{\cosh^2(x)}\,
  \Gamma\left(4,\frac{m_0}{\Tf}\cosh(x)\right) \,,
\\
\label{eq:F-omega}
  F_\omega(x) =\frac{ \tanh(x)}{\cosh^2(x)}\,
  \Gamma\left(4,\frac{m_0}{\Tf}\cosh(x)\right) \,,
\end{eqnarray}
while $\Corr_{XY}(\xi_1-\xi_2;\tauf)=\la X(\xi_1,\tauf) Y(\xi_2,\tauf)\ra$ is the equal-time 
rapidity correlator defined in Eq.~(\ref{eq:Corr-def}).
It is convenient to use Fourier transforms of those functions in terms of which
\begin{equation}
  \label{eq:dNdeta-dNdeta-k}
  \la \delta\frac{d N}{d\eta_1}\,\delta\frac{d N}{d\eta_2}\ra
=
\left(\frac{\ds \aperp\tauf \Tf^3}{4\pi^2}\right)^2\,
\int \frac{dk}{2\pi} e^{ik\Delta\eta} \sum_{\substack{X=\rho,\omega\\Y=\rho,\omega}}
\tilde F_X(-k) \tilde F_Y(k) \tilde \Corr_{XY}(k;\tauf),
\end{equation}
where $\Delta\eta=\eta_1-\eta_2$. 

\subsection{Normalization}
\label{sec:norm}

Since $\Corr_{XY}\sim 1/\aperp$ (see Eq. (\ref{eq:XY-k})), it is convenient to divide by the event average of $dN/d\eta$ given by Eq. (\ref{eq:dNdeta-A}) with
$\omega=0$ in order to remove the dependence on the transverse area $\aperp$ of the system.  Integrating over $d^2p_\perp$ we get
\begin{equation}
  \label{eq:dNdeta-ave}
  \la \frac{dN}{d\eta}\ra 
= \frac{\ds \aperp\tauf \Tf^3}{4\pi^2}
\int\!\frac{ dx}{\cosh^2(x)} \,
\Gamma\left(3,\frac{m_0}{\Tf}\cosh(x)\right)
\end{equation}
Collecting all factors in front of the integrals in Eqs. (\ref{eq:Corr-def}), (\ref{eq:dNdeta-dNdeta-k}) and (\ref{eq:dNdeta-ave}), 
one can write the normalized correlator as
\begin{equation}
  \label{eq:dNdN/N}
  \la \delta\frac{d N}{d\eta_1}\,\delta\frac{d N}{d\eta_2}\ra\la
  \frac{dN}{d\eta}\ra^{-1} = 
\frac{45\ds}{4\pi^4 N_{\rm eff}(T_0)}\,\frac{ \nu}{ \Tf\tauf}
\pfrac{T_0^2}{\Tf^2}^{\vs^{\!-2}-2} K(\Delta\eta) \,,
\end{equation}
where we defined the effective number of bosonic species such that $s(T)=2\pi^2N_{\rm eff}(T)\,T^3/45$ and $T_0\equiv T(\tau_0)$.  We also defined the dimensionless function $K(\Delta\eta)$ in such a way that most of the dependence on $T_0$ for the long-range tail is in the prefactor (using the observations at the end of Sec. \ref{sec:wake}).  

\begin{figure}[ht]
  \centering  
\includegraphics[width=0.6\linewidth]{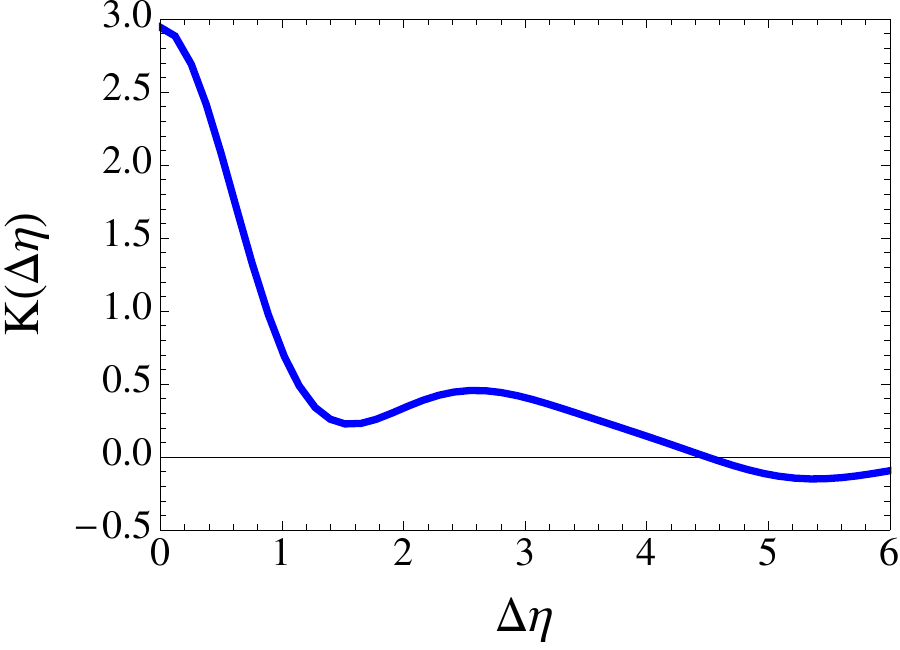}
  \caption[]{The correlation function $K(\Delta\eta)$ in the
    normalized correlator of $dN/d\eta$ fluctuations  in
    Eq. (\ref{eq:dNdN/N-ini}).  Viscosity is not included.}
\label{fig:thermonly}
\end{figure}

\begin{figure}[ht]
  \centering  
\includegraphics[width=0.6\linewidth]{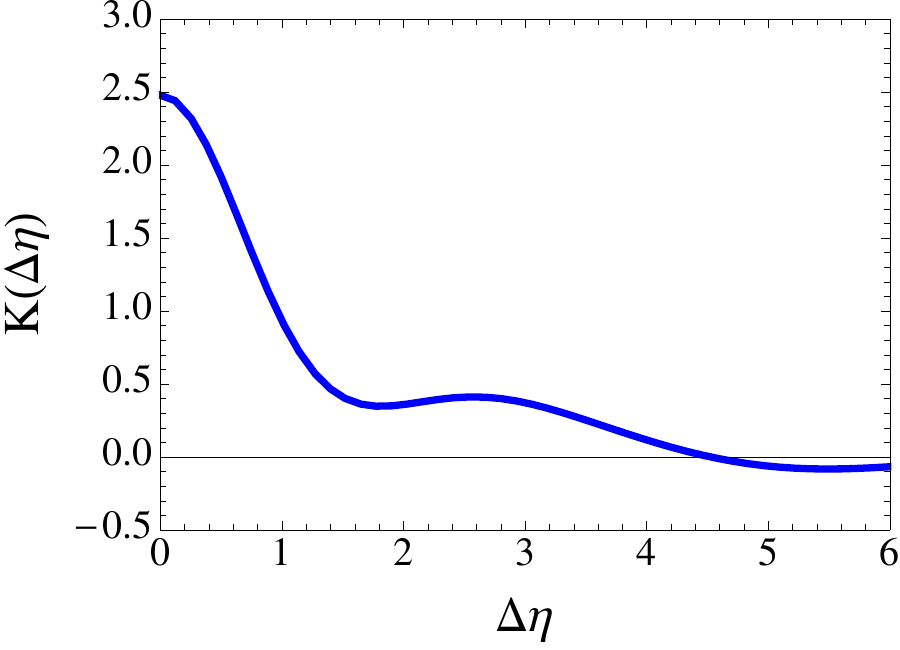}
  \caption[]{The correlation function $K(\Delta\eta)$ in the
    normalized correlator of $dN/d\eta$ fluctuations  in Eq. (\ref{eq:dNdN/N-ini}). Viscosity is included.}
\label{fig:visc}
\end{figure}

\begin{figure}[ht]
  \centering  
\includegraphics[width=0.6\linewidth]{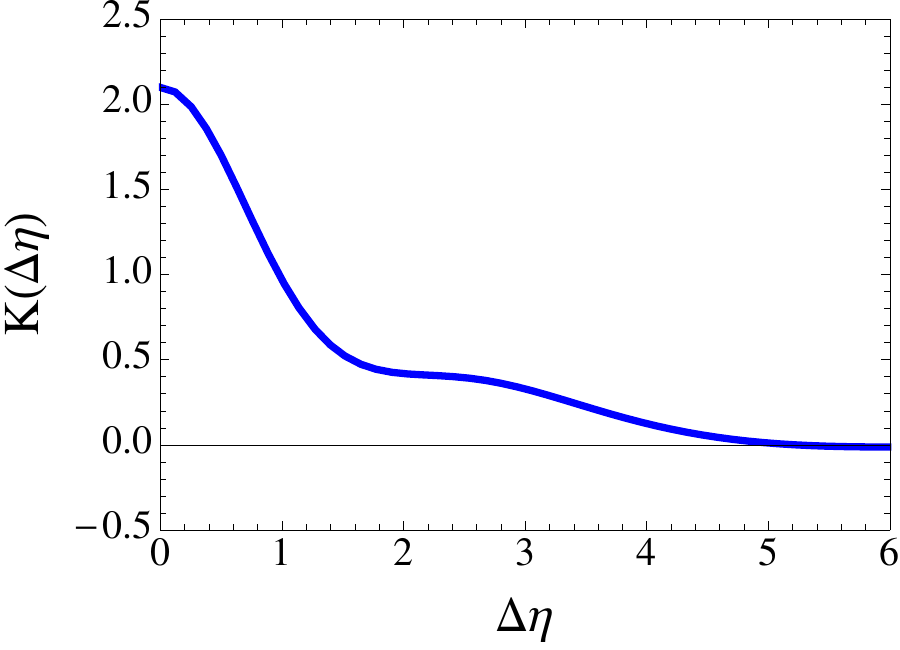}
  \caption[]{The correlation function $K(\Delta\eta)$ in the
    normalized correlator of $dN/d\eta$ fluctuations  in
    Eq. (\ref{eq:dNdN/N-ini}). Viscosity is included, as well as the 
source correlator broadening discussed in
    Sec. \ref{sec:noise-broadening}.}
\label{fig:expvisc}
\end{figure}

Assuming that reasonable values for the parameters are $\vs^2=1/3$, $\tauf=10$ fm, $\Tf=150$ MeV, $T_0=600$ MeV, $N_{\rm eff}=47.5$ (counting gluons and quarks) and $\nu=1/3\pi$, we plot the normalized correlator (\ref{eq:dNdN/N}) in Figs. \ref{fig:thermonly} and \ref{fig:visc}. To determine $K(\Delta\eta)$ we apply the freeze-out (thermal) smearing described by Eq. (\ref{eq:dNdeta-dNdeta-k}) in both plots under the assumption that the observed particles are pions. To evaluate the effect of viscosity, we compare  Fig. \ref{fig:thermonly}, which neglects viscosity, with Fig. \ref{fig:visc} which includes viscous broadening as described by Eqs. (\ref{eq:HT1T2}) and (\ref{eq:U2}).  The factor in front of $K(\Delta\eta)$ in Eq. (\ref{eq:dNdN/N}) is approximately $1.1\times 10^{-3}$ for our choice of parameters. Combining this with Fig. \ref{fig:visc}, we conclude that Eq. (\ref{eq:dNdN/N}) predicts correlations of the
order of $10^{-3}$.
Finally, to estimate the effect of higher-order hydrodynamic corrections we consider the noise correlator with non-zero correlation length as discussed in Sec. \ref{sec:noise-broadening}. Adding this effect on top of viscous broadening we obtain Fig. \ref{fig:expvisc}.  The effect is visible but does not
change the main features.

The important conclusion is that the absolute magnitude of the correlation (outside of the $|\xi|<1$ peak), for given freeze-out parameters, is proportional
to the relative viscosity $\nu$ and to a power of the initial temperature $T_0$.

\subsection{Contribution of initial state fluctuations}
\label{sec:contr-init-state}

Since the fluctuation equations (\ref{eq:dtpsi}) are linear and the noise is uncorrelated with initial conditions, the contribution of the initial state fluctuations to the correlator Eq. (\ref{eq:Corr-def}) is additive and is given by
\begin{equation}
  \label{eq:Corr-ini}
    \tilde\Corr_{XY}(k;\tauf)^{\rm ini}
=\sum_{X'Y'}
\tilde U_{XX'}(k;\tau,\tau_0)
\tilde\Corr_{X'Y'}(k;\tau_0)
\tilde U_{YY'}(-k;\tau,\tau_0)
\end{equation}
These correlations, unlike the purely hydrodynamic correlations discussed so far, depend also on the physics determining the initial-time correlator $\Corr_{XY}(\xi;\tau_0)$. Calculation of $\Corr_{XY}(\xi;\tau_0)$ is beyond hydrodynamics; it could be done from the traditional Glauber approach or from the color glass condensate (CGC) description of the initial state. However, once the initial correlator $\Corr_{XY}(\xi;\tau_0)$ is given, the subsequent evolution is governed by hydrodynamics according to Eq. (\ref{eq:Corr-ini}). Since hydrodynamic evolution is the main subject of this paper, we shall assume a generic form of the initial correlator $\Corr_{XY}(\xi;\tau_0)$, leaving its calculation beyond the scope of the paper. It is reasonable to assume that this correlator is local, meaning that $\tilde\Corr_{XY}(k;\tau_0)$ is a polynomial in $k$. We shall also assume, for simplicity, the following matrix form for it which obeys the basic symmetry properties of the correlator
\begin{equation}
  \label{eq:Corr-ini-def}
  {\tilde \Corr}_{XY}(k;\tau_0) 
\equiv  \la X(k;\tau_0)\,Y(-k;\tau_0)\ra 
=\frac{\cz}{\aperp}
\begin{pmatrix}
  1&-ik\\ik&k^2
\end{pmatrix}_{XY}
\end{equation}
where the factor $1/\aperp$  is due to the locality of the correlator in the transverse space (Eq. (\ref{eq:XY-k})) and where we defined the dimensionful coefficient $\cz$ which parameterizes the absolute strength of the correlator. Substitution into Eq. (\ref{eq:Corr-ini}) results in
\begin{equation}
  \label{eq:C-ini-G}
  \Corr_{XY}(k;\tau)^{\rm ini}=
\frac{\cz}{\aperp} G_{XY}(k;\tau,\tau_0) \,,
\end{equation}
where we used the definition of $G_{XY}$ in Eqs. (\ref{eq:G-GG}) and (\ref{eq:G-rho-U}).  Up to the constant factor $\cz$, the Fourier transform of this
correlator has been already discussed and its matrix element $G_{\rho\rho}$ has been plotted in Sec. \ref{sec:wake}. 

\begin{figure}[t]
  \centering  
\includegraphics[width=0.6\linewidth]{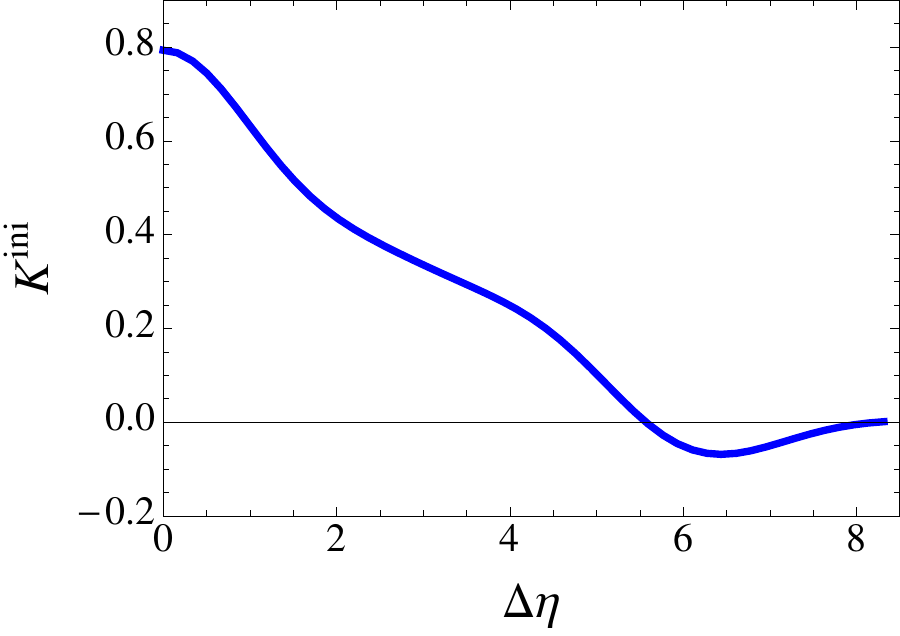}
\caption[]{The correlation function $K^{\rm ini}(\Delta\eta)$ in the
normalized correlator in Eq. (\ref{eq:dNdN/N-ini}). Viscosity is included.}
\label{fig:plot-ini}
\end{figure}

The contribution of such initial state fluctuations to the two-particle rapidity correlation, similarly to Eq. (\ref{eq:dNdeta-dNdeta-k}), is given by
\begin{equation}
  \label{eq:dNdeta-dNdeta-k-ini}
  \la \delta\frac{d N}{d\eta_1}\,\delta\frac{d N}{d\eta_2}\ra^{\rm ini}=
\cz \aperp\,\pfrac{\ds \tauf T^3}{4\pi^2}^2\,
\int \frac{dk}{2\pi} e^{ik\Delta\eta} \sum_{\substack{X=\rho,\omega\\Y=\rho,\omega}}
 \tilde F_X(-k) \tilde F_Y(k)  \tilde G_{XY}(k;\tauf,\tau_0) \,.
\end{equation}
Collecting the factors in front of the integrals in Eqs. (\ref{eq:dNdeta-dNdeta-k-ini}) and (\ref{eq:dNdeta-ave}) we can
write the contribution of initial state fluctuations to the normalized correlator as
\begin{equation}
    \label{eq:dNdN/N-ini}
  \la \delta\frac{d N}{d\eta_1}\,\delta\frac{d N}{d\eta_2}\ra^{\rm ini}
\la  \frac{dN}{d\eta}\ra^{-1} 
= \cz\,\frac{\ds \tauf \Tf^3}{4\pi^2}\,
 K^{\rm ini}(\Delta\eta)\,.
\end{equation}
Here we defined the function $K^{\rm ini}$ which we plot in Fig. \ref{fig:plot-ini} for the same choice of the parameters as in
Sec. \ref{sec:norm}. The coefficient in front of $K^{\rm ini}(\Delta\eta)$ in Eq.~(\ref{eq:dNdeta-dNdeta-k-ini}) is 
$8.5\times 10^{-3} (\cz/1\, {\rm GeV}^2)$  for that choice of parameters. Assuming that CGC initial conditions give rise to $\cz$ of the order
characteristic saturation scale  $c_0\sim Q_{\rm sat}^2\sim 1\,{\rm GeV}^2$ we conclude that such initial state fluctuations produce correlations of similar magnitude  to those due to purely hydrodynamic fluctuations. Of course, a more detailed analysis of initial conditions  is needed before a quantitative comparison with experiment  can be made.

\section{Summary and conclusions}
\label{sec:conclusion}

In this paper we explored the contribution of local hydrodynamic fluctuations to the event-by-event fluctuations of particles emitted from relativistic heavy ion collisions. Unlike the contribution of initial state fluctuations, which have been the main focus of the studies so far and whose magnitude is determined by quantum pre-equilibrium dynamics, the magnitude of the fluctuations we discussed here is directly related to the hydrodynamic properties of the locally equilibrated matter.  In the framework of relativistic viscous hydrodynamics, owing to the fluctuation-dissipation theorem, the amplitude of these fluctuations is governed by the viscosities. This offers a possibility to measure, or constrain, the viscosity of the strongly coupled quark-gluon plasma independent from the traditional analysis of elliptic flow.

We observed two remarkable features of the fluctuation correlator.  The first is that the correlations spread in rapidity space  logarithmically with Bjorken proper time, with velocity determined by the speed of the sound in a static medium. This behavior is  similar to the circles observed in \cite{Staig:2011as} for correlations in transverse space induced by initial state fluctuations.  The second is that we find the correlations are not limited to the sound cone but are accompanied by a wake behind the sound front which can be traced to the non-linearity of the sound mode dispersion relation in the medium. This diffusion-like wake is a non-dissipative process.

At the lowest order in the gradient expansion, the source of the fluctuations is white noise, which na\"ively leads to singularities in the correlation functions of the density and of the flow velocity. The singular behavior is tamed, in part, by the viscous terms in the hydrodynamic equations, as well as by the thermal smearing when one calculates final-state particle distributions. For completeness, we also considered the effect of replacing the white noise by colored noise with a thermal correlation function.

We explored the phenomenological consequences of hydrodynamic fluctuations in the idealized scenario of boost-invariant longitudinal flow with a homogeneous transverse profile.  In the Cooper-Frye approach to freeze-out, the correlation  function of the particle yield $dN/d\eta$ as a  function of the kinematic rapidity difference $\Delta\eta$ is obtained from the temperature and flow velocity correlation functions in Bjorken space rapidity $\xi$ by a thermal smearing.

Two features of the particle number correlation function deserve
special mention. One is a strong peak at $\Delta\eta = 0$, which
receives contributions from hydrodynamic fluctuations during the
entire course of the expansion. Its height and width is influenced by
both the thermal smearing at freeze-out and the viscous smearing
during the hydrodynamic phase (see also~\cite{Gavin-Aziz}). 

The other noteworthy feature is a broad structure at larger rapidity differences, extending up to the sound horizon $\Delta\eta_{\rm max} = 2 {\vs} \ln({\tauf}/\tau_0)$, which is caused by the hydrodynamic propagation of the noise followed by a slower diffusive wake generated at early times until thermal 
freeze-out. Depending on the precise values of the sound velocity and the start and end of the hydrodynamic phase, this implies that particle number correlations
extending over significantly more than one unit of rapidity can be generated during the hydrodynamic phase.  Correlations over large rapidity intervals have been observed in heavy-ion experiments \cite{Adams:2005ph,Putschke:2007mi} and have been subject to numerous theoretical studies \cite{Voloshin:2003ud,Armesto:2004pt,Strickland:2005we,Majumder:2006wi,Shuryak:2007fu,Wong:2007mf,Dumitru:2008wn,Chiu:2008ht}. It would be interesting to investigate to what extent the hydrodynamic fluctuations contribute to this phenomenon.

Let us emphasize again the main point of this paper.  It has been clear for some time that the profile of the particle number
correlations depends on the values of the shear (and bulk) viscosity
and of the sound velocity. As others have already argued
\cite{Staig:2010pn,Mocsy:2011xx} this is true for azimuthal
correlations generated by fluctuations in the hydrodynamic initial
conditions over the transverse plane.  Our results confirm this for
the longitudinal space correlations. What distinguishes hydrodynamic correlations induced by local thermal fluctuations is that their
absolute magnitude is also determined by hydrodynamic properties of
the medium. Thus our findings amplify the opportunities offered by  measured particle number fluctuations to constrain the fluid dynamical properties of the hot matter created in relativistic heavy ion collisions.

\acknowledgments

The work of J. K. was supported by the U.S. DOE Grant
No. DE-FG02-87ER40328, the work of B. M. by the U.S. DOE Grant
No. DE-FG02-05ER41367, and the work of M. S. by the U.S. DOE Grant
No. DE-FG02-01ER41195. We acknowledge the stimulating environment of
the Workshop on ``New Results from LHC and RHIC'' (embedded into the
INT-2a program) at the Institute for Nuclear Theory at the University
of Washington, where the research reported here was conceived.

\appendix

\section{Long-time limit}
\label{sec:long-time-limit}

Here we shall determine the behavior of the correlator $G_{\rho\rho}(\xi;\tauf,\tau)$ in the limit of asymptotically large time separation $\tauf/\tau\gg1$. We shall be able to obtain a closed analytical expression for the correlator in this limit. Unfortunately, for realistic values of $ \tauf/\tau$ relevant for heavy-ion collisions this
expression is still a poor approximation. However it is still useful, as analytic solutions often are, in demonstrating conceptually important features of the fluctuation correlator.

In the limit we consider the behavior of the correlator is determined by the modes with the slowest rate given by eigenvalues (\ref{eq:lambda12}). We observe that the slowest mode is $\lambda_-(k)$ at small $k$. This mode relaxes arbitrarily slowly as $k\to0$ like
\begin{equation}
  \label{eq:lambda-k-small}
  \lambda_-= \vs^2 \gamma_s^2 k^2+{\cal O}(k^4) \,,
\end{equation}
where $\gamma_s=1/\sqrt{1-v_s^2}$, while the mode $\lambda_+=1-\vs^2+{\cal O}(k^2)$. Thus in the limit $\tau/\tau'\gg1$ the longest lingering modes are smooth modes corresponding to the eigenvalue $\lambda_-$ given by $\bm{\tilde\psi_-}$ in Eq. (\ref{eq:psi-pm}). For such a mode
\begin{equation}
  \label{eq:omega-rho-lambda1}
  \tilde\omega=\frac{\lambda_-}{ik}\tilde\rho\approx -ik v_s^2 \gamma_s^2\tilde\rho \,.
\end{equation}
The matrix $\bm U$ in this limit becomes a projector on this mode, as can be seen directly from Eq. (\ref{eq:U-matrix}). Instead of proceeding from there, we
can also simply substitute Eq. (\ref{eq:omega-rho-lambda1}) into Eq. (\ref{eq:rho-eq}) and obtain a single equation for $\rho$
\begin{equation}
  \label{eq:2}
  \tau \frac{\pd\tilde\rho}{\pd\tau} + \lambda_- \tilde\rho +  \tilde{f} = 0
\end{equation}
which is easily solved in the form of Eq. (\ref{eq:rho-G-f}) with the Green's function $\tilde G_\rho(k;\tau,\tau')$ given by
\begin{equation}
  \label{eq:G-rho-rho}
 \tilde G_{\rho}(k;\tau,\tau') 
= \left({\tau'}/{\tau}\right)^{\lambda_- } \,.
\end{equation} 
Using equation~(\ref{eq:G-GG}) we find
\begin{equation}
  \label{eq:G-rho-rho-long}
 \tilde G_{\rho\rho}(k;\tauf,\tau) 
= \left({\tau}/{\tauf}\right)^{2\vs^2 \gamma_s^2 k^2}
\end{equation}
where we also used (\ref{eq:lambda-k-small}).  Fourier transforming from $k$ to $\xi$ we find
\begin{equation}
  \label{eq:G-tau-tau-xi}
  {G}_{\rho\rho}(\xi;\tauf,\tau) = \left( 8\pi \vs^2 \gamma_s^2 \ln(\tauf/\tau)\right)^{-1/2}
 \exp\left[-\frac{1}{8\vs^2 \gamma_s^2}\frac{\xi^2}{\ln(\tauf/\tau)}\right]
\end{equation}
We see that the fluctuation correlator describes diffusion in the Bjorken coordinate $\xi$ with $\ln(\tauf/\tau)$ playing the role of time. It is interesting that this diffusion-like process is not associated with dissipation.

Another question which can be asked is what happens to the singularities we discussed in Sec. \ref{sec:sing-sound-horiz} in the long-time limit. This can be easily seen by comparing Eq. (\ref{eq:G-tau-tau-xi}) with Eq. (\ref{eq:G-rhorho-sing}). One can see that the strength of the singularities decreases exponentially with
$\ln(\tauf/\tau)$ as $(\tau/\tauf)^{1-\vs^2}$, while the contribution of the smooth modes is roughly time-independent. In particular, the sum rule (\ref{eq:sr}) is
completely saturated by the Gaussian (\ref{eq:G-tau-tau-xi}) at late times. Numerically, at finite $\ln(\tauf/\tau)=4$ the regular part $G_{\rho\rho}^{\rm reg}$
in Fig.~\ref{fig:G_reg} contributes $1.12$, with the singular part $ G_{\rho\rho}^{\rm sing}$ accounting for  $-0.12$.

\section{Azimuthal correlations and power spectrum}
\label{sec:azim}

Our example application of the theory of hydrodynamic fluctuations, the one-dimensional boost invariant Bjorken flow, allowed us to study correlations of the hydrodynamic quantities in the longitudinal direction. There are several natural extensions. Here we shall make a few comments concerning the study of hydrodynamic correlations in the transverse plane. The natural quantity to consider in this case are azimuthal correlations among the number of emitted particles event by
event.

In contrast to initial state fluctuations, which are intimately related to geometric aspects of the nuclear collision, hydrodynamical fluctuations are caused by local noise and are thus unaffected by the global geometry. This suggests that it makes little sense to look for correlations between hydrodynamical fluctuations and global event properties, such as the event plane defined by the impact parameter or the elliptic flow pattern. Instead, a more promising analysis will follow the procedure used to determine the power spectrum of fluctuations in the cosmic microwave background radiation (CMBR) \cite{Dodelson}. The general argument for the application of this approach to heavy ion collisions has been proposed by Mishra {\em et al.} \cite{Mishra:2007tw,Mishra:2008dm} for the elliptic flow velocity $v_2$. 

The experimental determination of the power spectrum of azimuthal fluctuations relies on the measurement of two-particle correlations in the final state. This observable was originally suggested as a method of measuring collective flow anisotropies that does not require the determination of the reaction plane \cite{Wang:1991qh,Ollitrault:1997di,Borghini:2001vi}.  However, as Mishra, {\em et al.} emphasized, the power spectrum $|v_n|^2$ of the azimuthal anisotropy of the distribution of emitted particles not only picks up the collective flow anisotropy, but also event-by-event fluctuations of the emission pattern, including density fluctuations and fluctuations of the collective flow velocity. 

Within a chosen rapidity window, which will depend on the masses of the particles involved, one can represent the distribution in some observable $O$ as
\be
O(\phi) = {\overline O} + \sum_{n \neq 0} o_n {\rm e}^{i n \phi} ,
\ee
where $\phi$ denotes the azimuthal angle. The window is fixed in the laboratory frame of reference, and its orientation does not vary from one collision to the next.  For the CMBR the averaging is performed over points in the sky.  In the case of relativistic heavy ion collisions the averaging is done over a large set of events in which the orientation reaction plane and the hydrodynamic noise change randomly from one event to another. Thus
\be
\langle o_n \rangle = 0 \,,
\qquad\qquad
\langle o_n o_{n^{\prime}}^* \rangle = \frac{1}{2} O_n^2 \delta_{n,\, n^{\prime}} \, ,
\ee
where $O_n$ characterizes the magnitude (power) of the $n^{\rm th}$ angular Fourier component of the fluctuations. The dependence of $O_n$ on $n$ may reveal hydrodynamic and thermodynamic properties of the expanding medium. For example, as has been already observed in the studies of correlations induced by
initial state fluctuations \cite{Staig:2011as}, viscosity suppresses higher harmonics $n$. It would be also interesting to consider the effect of the QCD critical point (see  Ref. \cite{Stephanov:2004wx} for a review) on the power spectrum. In this case the natural choice of the variable $O$ would be the baryon density, or its experimental proxy -- net proton density. The magnitude of fluctuations increases near the critical point \cite{Stephanov:1999zu,Hatta:2003wn}, and the fluctuations can become highly non-Gaussian \cite{Stephanov:2008qz,Kapusta:2010ke}. The increase of fluctuations could be observed in the power spectrum. In addition, the increase in the correlation length may cause the power spectrum to shift its weight towards smaller values of $n$. Further quantitative investigation is needed, of course, to determine whether these effects have an observable magnitude. We leave this to future work.

\end{document}